\documentclass[%
 reprint,
 amsmath,amssymb,
 aps,
]{revtex4-2}

\usepackage{graphicx}
\usepackage{dcolumn}
\usepackage{bm}
\usepackage{tikz}
\usepackage{mathrsfs}  
\usepackage{slashed}
 \def\be{\begin{equation}}
 \def\ee{\end{equation}}
 \def\bea{\begin{eqnarray}}
 \def\eea{\end{eqnarray}}

\newcommand{\unit}{\mathbb{I}}

\begin{document}

\preprint{APS/123-QED}

\title{Shear viscosity and electric conductivity of  a hot and dense QGP with a chiral phase transition}
       \author{Olga Soloveva$^1$, David Fuseau$^2$, J\"org Aichelin$^2$, Elena Bratkovskaya$^{1,3,4}$}
		\email{soloveva@itp.uni-frankfurt.de}
		\address{$^1$ Institut f\"ur Theoretische Physik, Johann Wolfgang Goethe-Universit\"at,
		Max-von-Laue-Str.\ 1, D-60438 Frankfurt am Main, Germany}
		\address{$^2$ SUBATECH,  University  of  Nantes,  IMT  Atlantique,IN2P3/CNRS  4  rue  Alfred  Kastler,  44307  Nantes  cedex  3,  France}
		\address{$^3$ GSI Helmholtzzentrum f\"ur Schwerionenforschung GmbH,
		Planckstrasse 1, D-64291 Darmstadt, Germany}
		\address{$^4$ Helmholtz Research Academy Hessen for FAIR (HFHF), GSI Helmholtz Center for Heavy Ion Physics, Campus Frankfurt, 60438 Frankfurt, Germany}
		
\date{\today}

\begin{abstract}
We calculate two transport coefficients -- the shear viscosity over entropy ratio 
$\eta/s$ and the ratio of the electric conductivity to the temperature $\sigma_0/T$ -- 
of strongly interacting  quark matter within the extended $N_f=3$ Polyakov 
Nambu-Jona-Lasinio (PNJL) model along the crossover transition line for moderate 
values of baryon chemical potential $0 \leq \mu_B \leq 0.9$ GeV as well as 
in the vicinity of the critical endpoint (CEP) and at large baryon chemical potential 
$\mu_B=1.2$ GeV, where the first-order phase transition takes place. 
The evaluation of the transport coefficients is performed on the basis 
of the effective Boltzmann equation in the relaxation time approximation. 
We employ two different methods for the calculation of the quark relaxation times:
i) using the averaged transition rate defined via thermal averaged  quark-quark 
and quark-antiquark PNJL cross sections 
and 
ii) using the 'weighted' thermal averaged  quark-quark and quark-antiquark PNJL cross sections.
The $\eta/s$ and  $\sigma_0/T$ transport coefficients have a similar temperature and 
chemical potential behavior when approaching the chiral phase transition for the 
both methods for the quark relaxation time, however, the differences grow with increasing
temperature.
We demonstrate the effect of the first-order phase transition and of the CEP on the transport coefficients in the deconfined QCD medium.
\end{abstract}

\maketitle

\section{Introduction}

Understanding  the nature of a possible  phase transition
and the properties of the quark-gluon plasma (QGP) produced in 
relativistic heavy-ion collisions  \cite{Shuryak09, Heinz13, Adronic18} is presently one of the most challenging questions in the physics of strong interactions. 
State-of-the-art lattice QCD (lQCD) calculations allow for the evaluation of 
the thermal properties of the QGP at vanishing baryon  chemical potential ($\mu_B$). 
For finite baryon chemical potential one has to rely on phenomenological models.
Moreover, for the calculation of transport coefficients one has to 
advance to  transport theories which describe the expansion of the QGP. 
The experimental exploration of the finite $\mu_B$ region of the QCD phase diagram
is one of the primary goals of the  Beam Energy Scan programs of RHIC at BNL \cite{Odyniec:2019kfh} as well as of the planned experimental program of FAIR 
(Facility for Antiproton and Ion Research) \cite{Senger:2020pzs} at GSI and of the NICA 
(Nuclotron-based Ion Collider fAcility) facility at JINR \cite{Blaschke:2016nica}. 

In a hot and dense environment it is notoriously difficult to calculate microscopic properties of the QGP from first principles \cite{AMY:eta2003}. The expansion and dilution of the quark medium, produced in nuclear collisions, is usually  described by relativistic viscous hydrodynamics, which contains transport coefficients in the dissipative part. 
Although the hydrodynamic equations provide a macroscopic description of a relativistic fluid behavior, transport coefficients are sensitive to the underlying microscopic theory. They provide  information about the interactions inside the medium.
The most frequently studied  transport coefficient is the shear viscosity, which is used in viscous hydrodynamic simulations \cite{Shuryak:2017zxa}. It has been shown that also the bulk viscosity plays an important role for the time evolution of the QGP \cite{Ryu}.

A growing number of studies have examined transport coefficients of the QGP on the basis of effective models at zero or small values of the chemical potential \cite{Attems:2016ugt,Rougemont:hologr,Jackson:2017hfz,Thakur,Greif18,Mykhaylova:2019wci,Zhao:2020xob,Abhishek:2020wjm} where lQCD calculations can serve as a guideline.
In order to extend the study of transport coefficients to the part of the phase diagram
where the phase transition is possibly changing from a crossover to a 1st order one it is necessary to resort to  effective models which describe the chiral phase transition. 

The goal of this study is to calculate two transport coefficients of the QGP --
the shear viscosity over entropy ratio 
$\eta/s$ and the ratio of the electric conductivity to the temperature $\sigma_0/T$ -- 
at finite temperature and chemical potential 
within the framework of an effective relativistic Boltzmann equation in the relaxation-time  approximation, where properties of the QGP matter such as the equation-of-state (EoS), the interaction cross sections and constituent quark masses are described by the extended Polyakov Nambu-Jona-Lasinio (PNJL) model with a critical endpoint CEP located at ($T^{CEP}$, 
$\mu^{CEP}_q) = (0.11, 0.32)$ GeV \cite{Fuseau:2019zld}. 
Our study is based on  the  advances  of a  previous  work \cite{Marty:NJL13}
where the transport coefficients have been calculated within the NJL model at $\mu_B=0$.
However, now we use an advanced Polyakov extension of the NJL model in which
the PNJL EoS equals the lQCD EoS at vanishing $\mu_B$  - cf. \cite{Fuseau:2019zld}. 
Moreover, the framework of the PNJL model allows to calculate the transport properties near the chiral phase transition at moderate and high $\mu_B$. Here we denote the  quark chemical potential as (for the light quarks) $\mu_q=\mu_l=\mu_B/3$ while for the strange quarks we take $\mu_s=0$. \\
We compare the PNJL transport coefficients with those from Ref. 
\cite{Moreau:2019vhw,Soloveva:2019xph}, where they have been calculated within the dynamical 
quasi-particle model (DQPM) \cite{Peshier:2005pp,Cassing:2007nb} 
at moderate values of the baryon chemical potential, $\mu_B \leq 0.5$ GeV.
The both models are based on rather different ideas: 
The DQPM is an effective model for the description of non-perturbative 
(strongly interacting) QCD based on the lQCD EoS.
The phase transition there is a crossover for zero as well as 
for finite $\mu_B$.
The degrees-of-freedom of the DQPM are strongly interacting dynamical quasiparticles --
quarks and gluons -- with broad spectral functions, whose 'thermal' masses and widths increase with growing temperature, 
while the degrees-of-freedom of the PNJL are quarks whose masses approach the bare
mass when the temperature increases and the chiral condensate disappears. 
Moreover, in the PNJL the mesons are existing above the Mott transition temperature as resonances.
Thus, in the EoS the energy density is shared between the quarks, mesons and the Polyakov loop potential. 
We  explore how the nature of the degrees-of-freedom affects the transport 
properties of the QGP.
Moreover, we study the possible influence of the presence of a CEP and of a 1st order phase transition at high baryon chemical potential. 
For the evaluation of the relaxation time we  use two 
different methods: a)  the 'averaged transition rate' defined via the thermal averaged  
quark-quark and quark-antiquark PNJL cross sections
 and  b)  the 'weighted' thermal averaged  quark-quark and quark-antiquark PNJL cross sections (here under 'thermal averaged' cross section we mean the averaging 
 of the  interaction cross section over the thermal (anti-)quark distribution function). 
We discuss the uncertainties related to the theoretical methods based on the RTA.

The paper is organized as follows: In Sec.\ref{sec2} we give a brief review of the basic ingredients of the PNJL model, detail the description of the evaluation of the total quark cross sections for different channels, and show the temperature and chemical potential dependence of the total cross sections. In Sec.\ref{sec3} we discuss the computation of  the specific shear viscosity and the electric conductivity based on the relaxation time approximation of the Boltzmann equation. We consider two methods for the evaluation of the quark relaxation times and discuss differences between them. We compare furthermore our results at $\mu_B=0$ to calculations from lQCD for $N_f=0$ for the specific shear viscosity and to lQCD results for $N_f=2$ and $N_f=2+1$ for the electric conductivity and to  predictions of the DQPM for both transport coefficients. In addition, we show the ratio of dimensionless transport coefficients for the full range of chemical potentials. We finalize our study with conclusions in Sec.\ref{sec4}.


\section{\label{sec2} PNJL quark-quark cross sections}
We start with the calculation of the quark-quark elastic cross sections at finite temperature and chemical potential using scalar and pseudoscalar mesons 
as exchanged boson within the PNJL model of Ref. \cite{Fuseau:2019zld}. This model is an improved version of the standard Polyakov extended Nambu-Jona-Lasinio model where the pressure is calculated next to leading order in $N_c$  and the effective potential is phenomenologically re-parametrised to describe a medium in which also quarks are present.
Correspondingly, the masses of the quarks and the mesonic propagators are evaluated using this upgraded PNJL model \cite{Fuseau:2019zld}. 
We note that the cross sections for quark-meson or quark-diquark channels 
can also be found in Ref. \cite{Friesen:2013bta}. They are not used in this study.

\subsection{PNJL model}
	
The PNJL \cite{Fukushima:2003fw,Megias:2004hj,Ratti:2005jh,Hansen:2006ee,Torres-Rincon:2015rma} model is an extension of the NJL model including thermal gluons on the level of a mean field. The quark-quark interaction remains local, the gluons being only present as an effective potential surrounding the quarks. It can be associated to the $\frac{1}{4}F_{\mu\nu}^aF^{a\mu\nu}$ term in the QCD Lagrangian.
The Lagrangian of the PNJL model~\cite{Fukushima:2003fw,Megias:2004hj,Ratti:2005jh,Hansen:2006ee,Torres-Rincon:2015rma} 
for (color neutral) pseudoscalar and scalar interactions (neglecting the vector and axial-vector vertices for simplicity) is
\begin{align}
	\mathscr{L}_{PNJL} &= \sum_i \bar{\psi}_i (i \slashed {D}-m_{0i}+\mu_{i} \gamma_0) \psi_i \nonumber \\
      &+ G \sum_{a} \sum_{ijkl} \left[ (\bar{\psi}_i \ i\gamma_5 \tau^{a}_{ij} \psi_j) \ (\bar{\psi}_k \ i \gamma_5 \tau^{a}_{kl} \psi_l)\right.\nonumber\\
&\quad +\left.(\bar{\psi}_i \tau^{a}_{ij} \psi_j) \ (\bar{\psi}_k  \tau^{a}_{kl} \psi_l) \right] \nonumber \\
& -    K \det_{ij} \left[ \bar{\psi}_i \ ( \unit - \gamma_5 ) \psi_j \right] - K \det_{ij} \left[ \bar{\psi}_i \ ( \unit + \gamma_5 ) \psi_j \right]  \nonumber \\ 
&- {\cal U} (T;\Phi,\bar{\Phi})\ .
\label{eq:lagPNJL}
\end{align}
Here $i,j,k,l=1,2,3$ are the flavor indices and $\tau^{a}$ ($a=1,...,8$) are the $N_f=3$ flavor generators with the normalization
\begin{equation}
\textrm{tr}_f \  (\tau^{a} \tau^{b}) = 2\delta^{ab}  \ ,
\end{equation}
with $\textrm{tr}_f$ denoting the trace in the flavor space. $m_{0i}$ stands for the bare quark masses and $\mu_{i}$ for their chemical potential. The covariant derivative in the Polyakov gauge reads $D^\mu=\partial^\mu - i \delta^{\mu 0} A^0$, with $A^0=-iA_4$ being the temporal component of the gluon field in
Euclidean space (we denote $A^\mu = g_s A_{a}^\mu T_{a}$). The coupling constant for the scalar and pseudoscalar interaction $G$ is taken as
a free parameter (fixed e.g. by the pion mass in vacuum). The value of the free parameters of the PNJL are displayed in table \ref{table PNJL0}.
  \begin{table}[h!]	    \centerline{\begin{tabular}{|c|c|c|c|c|}
	    \hline
	    $m_u[GeV]$ & $m_s[GeV]$ & G & K & $\Lambda [GeV] $ \\
	    \hline
	    0.005 & 0.134 & $\frac{2.3}{\Lambda^2}$ & $\frac{11.}{\Lambda^5}$ & 0.569\\
	    \hline
	    \end{tabular}}
	    \caption{Table of the parameters of the PNJL model used in this paper.}
	    	    \label{table PNJL0}
	      \end{table}
The third term of Eq.~(\ref{eq:lagPNJL}) is the so-called 't Hooft Lagrangian which makes the mass splitting between the $\eta$ and the $\eta'$ mesons known as the axial $U(1)$ anomaly. $K$ is a coupling constant (fixed by the value of $m_{\eta'}-m_{\eta}$) and $\unit$ is the identity matrix in Dirac space.

 Finally, ${\cal U} (T,\Phi,\bar{\Phi})$ is the so-called Polyakov-loop effective potential used to account for the static gluonic contributions to the pressure. The Polyakov line and the Polyakov loop are, respectively, defined as
 \begin{equation}
 	 L({\bf x}) = {\cal P} \exp \left( i \int_0^{1/T} d\tau A_4 (\tau,{\bf x}) \right)
 	 \end{equation}
 	 and
 	 \begin{equation}
\Phi ({\bf x})= \frac{1}{N_c} {\rm tr}_c L({\bf x}) \ ,
 \end{equation}
where ${\cal P}$ is the path-integral ordering operator, and the trace ${\rm tr}_c$ is taken in the color space.

The value of the potential for the expectation values $\langle \Phi \rangle (T),\langle \bar{\Phi} \rangle (T)$ ,  $ {\cal U} (T, \langle \Phi \rangle (T), \langle \bar{\Phi} \rangle (T))$,
gives up to a minus sign the pressure of the gluons in Yang Mills (YM) theory, corresponding to QCD for infinitely heavy quarks. The comparison with lattice gauge calculations 
for pure YM serves therefore as a guideline for the parametrisation of the effective potential $U(T)$
\begin{equation}
	-P(T) ={\cal U} (T, \langle \Phi \rangle (T), \langle \bar{\Phi} \rangle (T)).
\end{equation}
	    
	    The effective potential $U(T,\phi=\langle \Phi \rangle (T), \bar{\phi}=\langle \bar{\Phi} \rangle (T))$ is parametrised following Ref.~\cite{Fuseau:2019zld}:    
    \begin{equation}
	    \frac{U(\phi, \bar{\phi},T)}{T^4} = -\frac{b_2(T)}{2}\bar{\phi}\phi - \frac{b_3}{6}(\bar{\phi}^3 + \phi^3) + \frac{b_4}{4}(\bar{\phi}\phi)^2
	    \label{Potentiel polynome Polyakov}
	    \end{equation}
	    with the parameters 
		\begin{equation}
		b_2(T) = a_0 + \frac{a_1}{1+\tau} + \frac{a_2}{(1+\tau)^2} + \frac{a_3}{(1+\tau)^3}
\end{equation}			 
 where
	    \begin{equation}
	    	t_{phen} = 0.57\frac{T-T_{phen}(T)}{T_{phen}(T)}
	    	 \label{Temperature reduite}
	    \end{equation}
  and
	        \begin{equation}
		  T_{phen}(T) = a + bT + cT^2 + dT^3 + e\frac{1}{T}.
		  \label{T0 poly}
		\end{equation}
    All the coefficients of the parametrisation are listed in table \ref{table PNJL}.
	    \begin{table}[h!]
\centerline{\begin{tabular}{|c|c|c|c|c|c|c|c|c|c|c|}
	    \hline
	    $a_0$ & $a_1$ & $a_2$ & $a_3$ & $b_3$ & $b_4$ & a&b&c&d&e\\
	    \hline
	    6.75 & -1.95 & 2.625 & -7.44 & 0.75 & 7.5 & 0.082&0.36& 0.72&-1.6&-0.0002\\
	    \hline
	    \end{tabular}}
	    \caption{Table of the parameters of the effective potential U($\phi$, $\bar{\phi}$,T) used in this paper.}
	    	    \label{table PNJL}
	      \end{table}
 \begin{center}
	\begin{figure*}
\begin{minipage}[h]{0.4\linewidth}
\center{\includegraphics[width=1\linewidth]{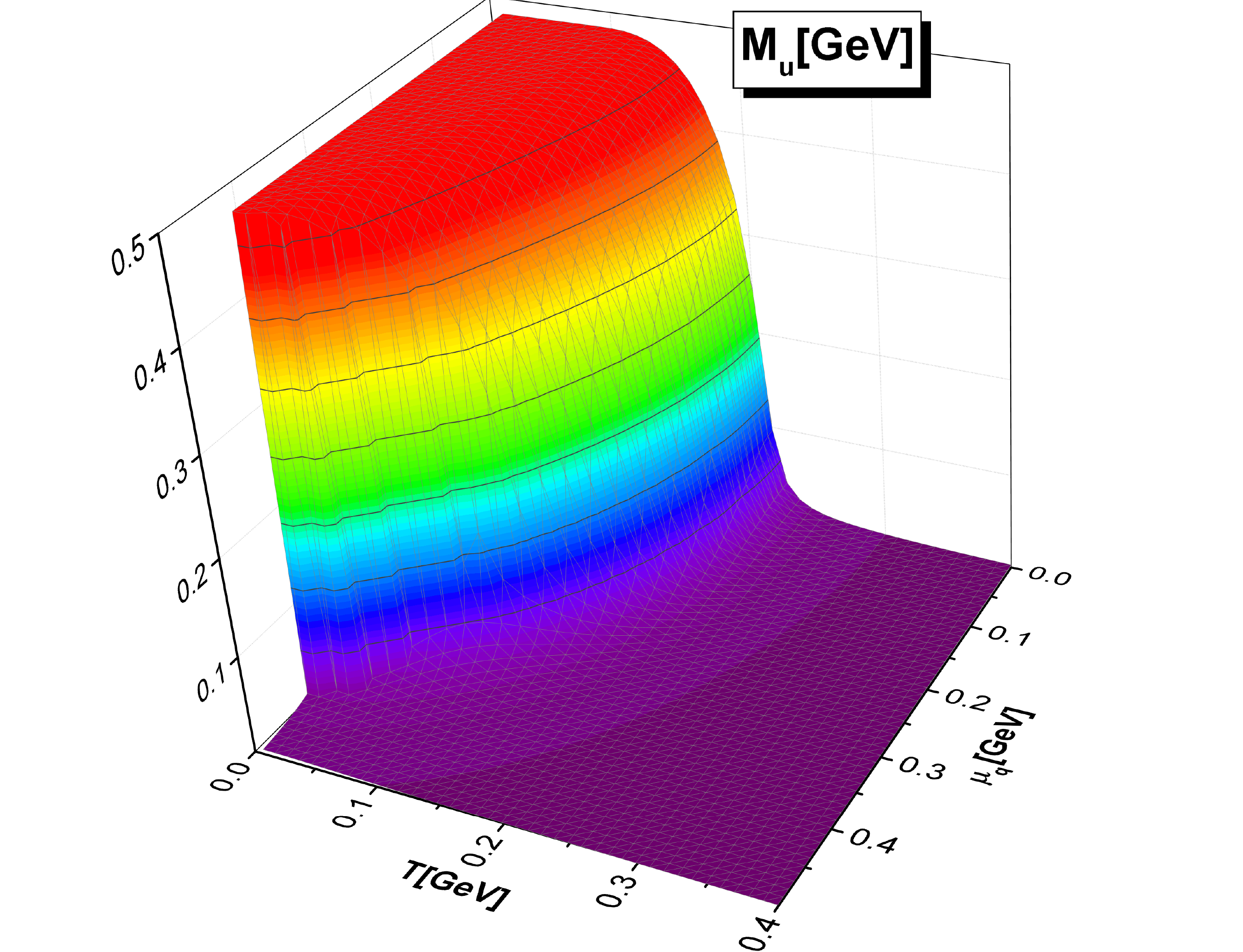} \\ a) light quark}
\end{minipage}
\begin{minipage}[h]{0.4\linewidth}
\center{\includegraphics[width=1\linewidth]{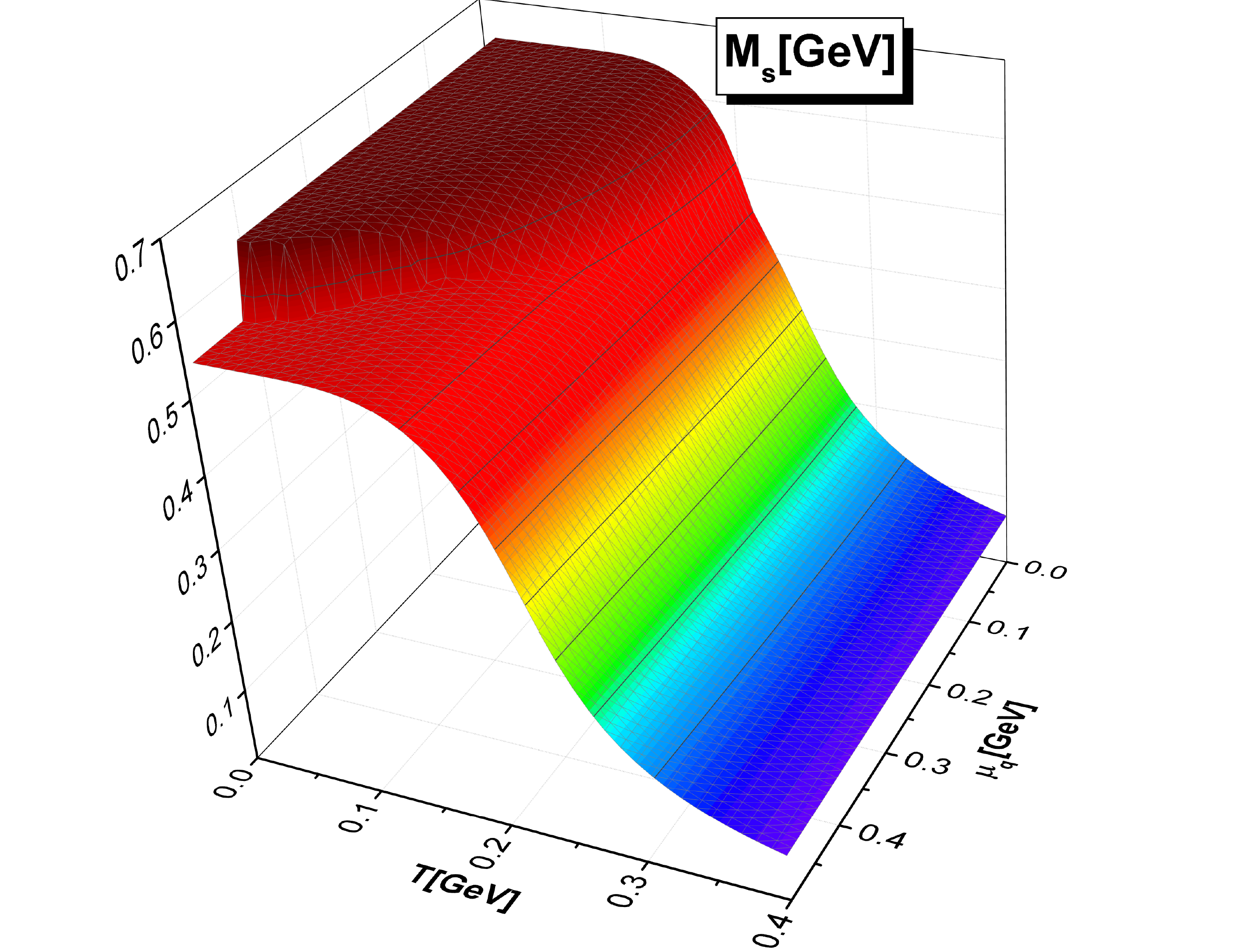} \\ b) strange quark}
\end{minipage}
\caption{ The masses of light (left) and strange (right) quarks 
as a function of temperature $T$ and quark chemical potential $\mu_q$ calculated
within the PNJL model. }
\label{masses}
    \end{figure*}
\end{center} 

    \subsection{Quarks and mesons in the PNJL}
    
	In order to calculate cross sections, the masses of the quarks and the propagators of the exchanged mesons have to be known.	
	The mass of the quarks is determined by solving the traditional gap equations \cite{Hansen:2006ee} together with a minimisation of the grand potential $\Omega_{PNJL}$ with respect to the Polyakov loop expectation value $\phi$ and $\bar{\phi}$:
	\begin{align}
	    &\frac{\partial\Omega_{PNJL}}{\partial\phi} = 0 \nonumber \\
	    &\frac{\partial\Omega_{PNJL}}{\partial\bar{\phi}} = 0\nonumber \\
	    m_q &= m_{q0} - 4 G \langle  \bar{\psi}_q \psi_q \rangle + 2 K  \langle \bar{\psi}_q \psi_q \rangle  \langle \bar{\psi}_s \psi_s  \rangle \ ,\nonumber\\
	    m_s &= m_{s0} - 4 G \langle  \bar{\psi}_s \psi_s \rangle + 2 K  \langle \bar{\psi}_q \psi_q \rangle  \langle \bar{\psi}_q \psi_q  \rangle \ ,
	     \label{gapeq}
	\end{align}
  where
 \begin{align}
  &\Omega_{PNJL} (T,\mu_i,\Phi,\bar{\Phi})=\nonumber\\
 & 2G \sum_i \langle \bar{\psi}_i \psi_i \rangle^2
 -4K \prod_i  \langle \bar{\psi}_i \psi_i \rangle  - 2N_c \sum_i \int \frac{d^3 k}{(2\pi)^3} E_i \nonumber \\
&- 2 T N_c \sum_i \int \frac{d^3 k}{(2\pi)^3} \left( \frac{1}{N_C}\log \right.\left[ 1 \right.\nonumber \\
&\left.+ 3(\Phi + \bar{\Phi} e^{-(E_i-\mu_i)/T})\right. 
\left. e^{-(E_i-\mu_i)/T}\right] \nonumber \\
  &+\frac{1}{N_C} \log \left[ 1 + 3(\bar{\Phi} + \Phi e^{-(E_i+\mu_i)/T}) e^{-(E_i+\mu_i)/T}\right]\left.\right)\nonumber\\
  &+ U_{PNJL} \ ,  
\label{OmegaMF}
\end{align}
with $E_i=\sqrt{k^2+m_i^2}$.\\
Fig.~\ref{masses} shows the masses of the u and s quarks as a function of the chemical potential $\mu_q$ and the temperature. One can see that a smooth crossover occurs for small chemical potentials. The chiral condensate $<\bar{\psi}\psi>$ goes from its maximal value in the hadronic phase down to zero in the QGP phase. At low temperature, this transition is discontinuous and a first-order phase transition occurs that ends with a critical endpoint(CEP) at $\mu_q=0.320$ GeV, $T=0.110$ GeV.

The propagators of the mesons are build by resummation of the quark-antiquark loops, leading to the amplitude \cite{Klevansky:1992qe}:
\begin{equation}
	\mathscr{D} = \frac{2ig_m}{1 - 2g_m\Pi_{ff'}^{\pm}(k_0,\vec{k}),}
	\label{amplimesonex}
\end{equation}
 where $g_m$ is the coupling constant \cite{Rehberg:1996vd} and $\Pi_{ff'}^{\pm}(k_0,\vec{k})$ is the polarisation function given by
   \begin{align}
	    \Pi_{ff'}^{\pm}(k_0,\vec{k}) = &-\frac{N_c}{4\pi^2}
	    \left[A_0(m_f,\mu_f,T,\Lambda) + A_0(m_{f'},\mu_{f'},T,\Lambda)\right.\nonumber\\
	    &+ [(m_f \pm m_{f'})^2 - (k_0 + \mu_f - \mu_{f'})^2 + \vec{k}^2]\nonumber\\
	    &\left. \times B_0(\vec{k},m,\mu,m',\mu',k_0,T,\Lambda)\right],
	    \label{Polarisation mesons ps}
    \end{align}
	  where  "+"  stands for the scalar and  "-" for the pseudoscalar mesons.
	
The one-fermion loop $A_0$ is separated into a vacuum part and a thermal part, the latter being integrated up to infinity:

    \bea
	      &	A_0(m_f,\mu_f,T,\Lambda) = -4(\int_0^\infty dp\frac{p^2}{E_f} \nonumber\\
	      	& \times \left[-f_f(E_f,T,\mu_f)-f_{\overline{f}}(E_f,T,\mu_f)\right]+ \int_0^\Lambda \frac{p^2dp}{E_f}   ),
	      \eea
	     
	      where $E_f=\sqrt{p^2 + m_f^2}$ and the Fermi-Dirac distribution functions $f_{f(\bar f)}(E_f,T,\mu_f)$  are defined as \\
	          	   	\begin{equation}
	f_{f(\bar f)}(E_f,T,\mu_f) = \frac{1}{e^{(E_f\pm \mu_f)/T}+1}. 
	   \end{equation}

The two-Fermion loop $B_0$ is defined as \cite{Rehberg:1995nr}

\bea
	      	&B_0&(\vec{p},m_f,\mu_f, m_{f'}, \mu_{f'}, i\nu_m,T,\Lambda)  \nonumber \\ 
                      &=&  16\pi^2T\sum_n\exp(i\omega_n\eta) \int_{|q|<\Lambda} \frac{d^3q}{(2\pi)^3}\nonumber \\
                     & \times&  \frac{1}{\left[(i\omega_n+\mu_f)^2-E_f^2\right]} \frac{1}{\left[(i\omega_n-i\nu_m+\mu_f')^2-E_{f'}^2\right]},
\eea
with $E_f = \sqrt{\vec{q}^2 + m_f^2}$, $E_{f'} = \sqrt{(\vec{q} - \vec{p})^2 + m_{f'}}$.
The details of the calculations of $B_0$ can be found in Ref.~\cite{Rehberg:1995nr}.

We have $g_m^\pm = G \pm \frac{1}{2}KS^s$ for the pion (+) and its scalar partner (-) and $g_m^\pm = G \pm \frac{1}{2}KS^u$ for the kaon (+) and its scalar partner (-).
For the $\eta$ mesons, the propagators are more complicated because of the mixing terms \cite{Rehberg:1995kh},\cite{Klevansky:1992qe}:
	\begin{equation}
		\mathscr{D} = 2\frac{det K}{M_{00}M_{88}-M_{08}^2}\begin{pmatrix}M_{00}&M_{08}\\ M_{80}&M_{88}\end{pmatrix}. 
	\end{equation}
	\begin{align}
		\mathscr{D}&= \frac{4}{3}\frac{det K}{M_{00}M_{88}-M_{08}^2}(M_{00}\bar{\psi}\lambda_0\psi \cdot \bar{\psi'}\lambda_0\psi'\nonumber\\ &+ M_{08}\bar{\psi}\lambda_0\psi \cdot\bar{\psi'}\lambda_8\psi'\nonumber\\ &+M_{80}\bar{\psi}\lambda_8\psi \cdot \bar{\psi'}\lambda_0\psi'\nonumber\\ &+ M_{88}\bar{\psi}\lambda_8\psi \cdot \bar{\psi'}\lambda_8\psi')
	\end{align}
	with
	\begin{align}
		M_{00} &= K_0^+ - \frac{4}{3}det K(\Pi_{u\bar{u}} + 2\Pi_{s\bar{s}})\\
		M_{08} &= K_{08}^+ - \frac{4}{3}\sqrt{2}det K(\Pi_{u\bar{u}} - \Pi_{s\bar{s}})\\
		M_{88} &= K_8^+ - \frac{4}{3}det K(2\Pi_{u\bar{u}} + \Pi_{s\bar{s}})\\
		det K &= K_0^+K_8^+ - K_{08}^2
	\end{align}
	and
	\begin{align}
	K_0^\pm &= G \mp \frac{1}{3}K(2 G^u + G^s)\\
	K_{08}^\pm &= \pm \frac{1}{6}\sqrt{2}K(G^u - G^s)\\
	K_8^\pm &= G \pm \frac{1}{6}K(4 G^u - G^s),
	\end{align}
where $G^i$ is the  spinor trace of the propagator $S^i(x,x)$:
\begin{equation}
    G^i = N_C i Tr[S^i(x,x)] = -\frac{N_C}{4\pi^2}m_iA_0(m_i,\mu_i,T,\Lambda).
\end{equation}

\begin{figure}[h!]
	 \centering
	 \includegraphics[scale=0.5]{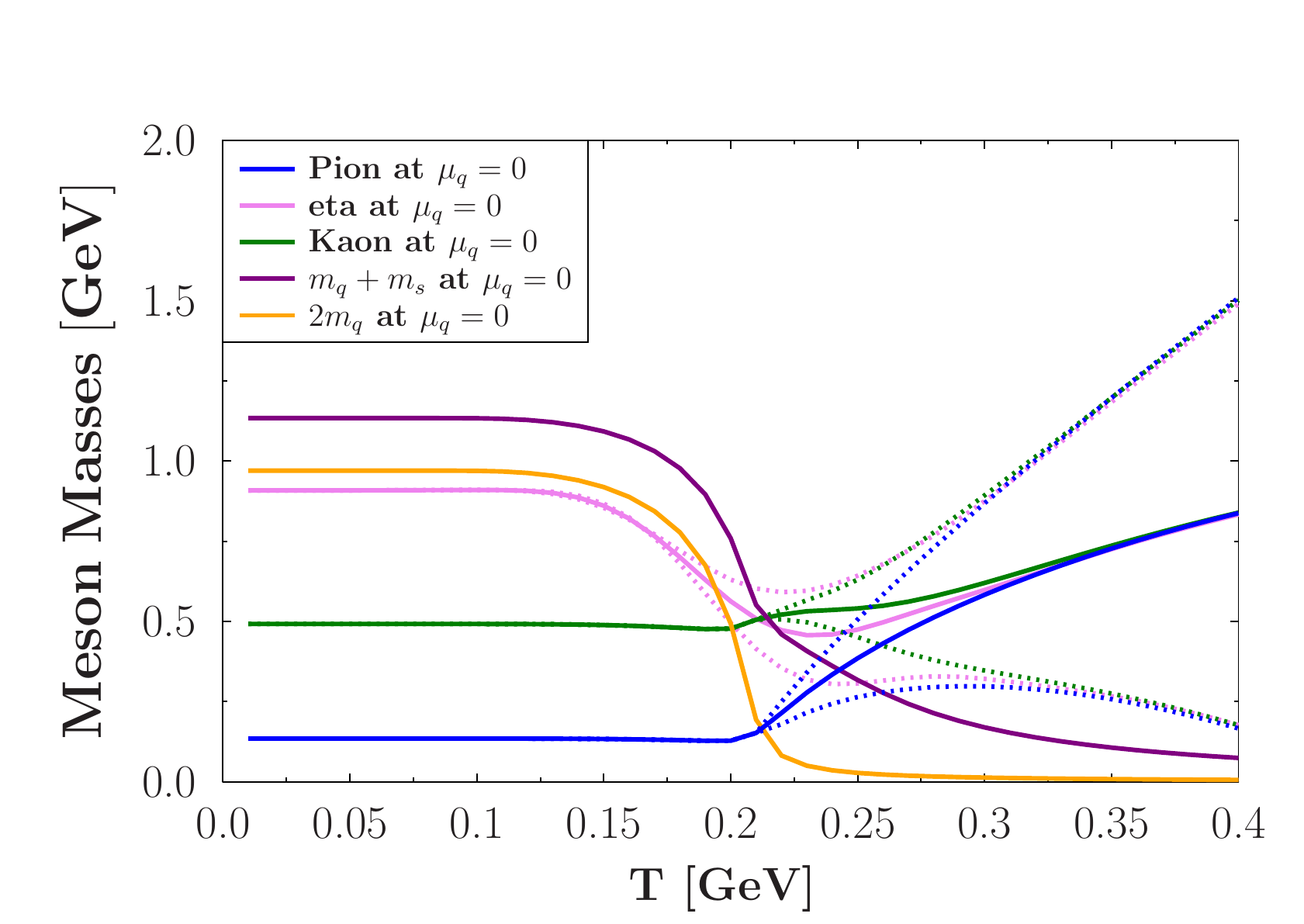}
	 \caption{ The PNJL results for the temperature dependence of the meson masses 
	 (pion, eta, kaon) as well as double quark masses $2m_q$ and $m_s+m_q$ for $\mu_q=0$.
  The dotted lines indicate the $m_{pole} \pm \Gamma$, where 
  the $\Gamma$ is the imaginary part of the complex pole of the meson propagators 
  and $m_{pole}$ is its real part, indicated by solid lines.}
	 	 \label{mes mas cs}
	 \end{figure}

The masses of the pseudoscalar mesons (pion, eta, kaon) at $\mu_q=0$
 are presented in Fig. \ref{mes mas cs}. The doubled quark masses $2m_q$ and $m_q+m_s$ 
 are shown for comparison, too. While in  PNJL the quark masses drop with increasing temperature 
to their bare values due to the disappearance of the chiral condensate in the vicinity of the phase transition,
 the meson masses increase with temperature. In PNJL  mesons become unstable
 above the Mott temperature, $T>T_{M\pi}$, where the total mass of the constituent quarks 
equals the meson mass. Above  $T_{M\pi}$ the mesons can decay into a $q\bar q$ pair. Therefore
the pole of the meson propagator becomes complex above $T_{M\pi}$ .
 The dotted lines in Fig. \ref{mes mas cs} indicate the $m_{pole} \pm \Gamma$, where 
 $\Gamma$ is the imaginary part of the complex pole of the meson propagator 
  (which could be associated  to the decay width)  and $m_{pole}$ is its real part, indicated by solid lines.
  Contrary to mesons, the quarks in the PNJL stay on-shell. 

	\subsection{The PNJL equation-of-state}

\begin{centering}
	    \begin{figure}[h!]
	    \centerline{\hspace*{-0.4in}\includegraphics[scale=0.5]{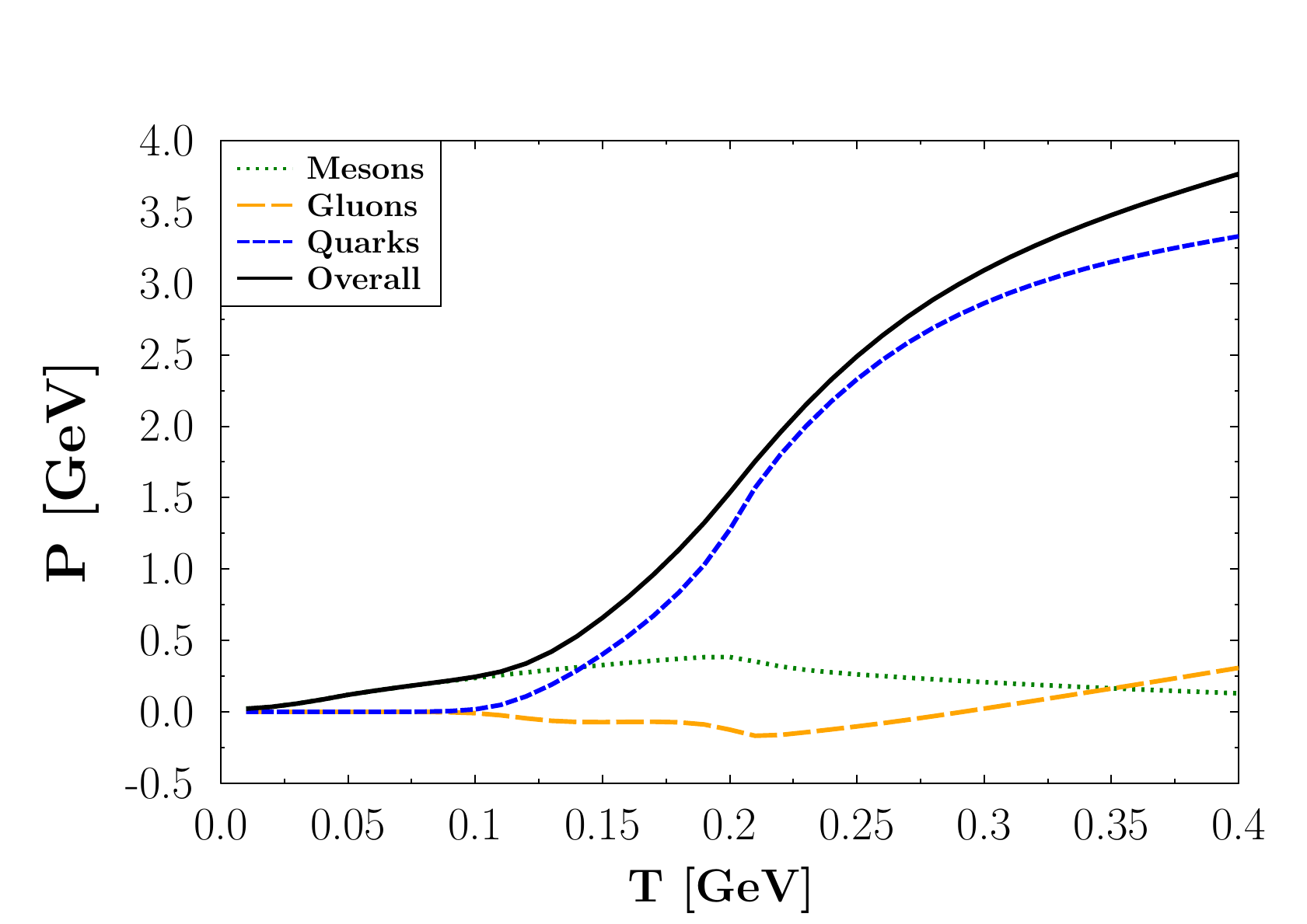}}
	    \caption{Mesons, gluons, quarks contributions to the total pressure 
	    as well as the total pressure (solid black line) at $\mu_q = 0$ as a function of the temperature.}
	    \label{PQ}
	    \end{figure}
	    \end{centering}

\begin{centering}
	    \begin{figure}[h!]
	    \centerline{\hspace*{-0.4in}\includegraphics[scale=0.5]{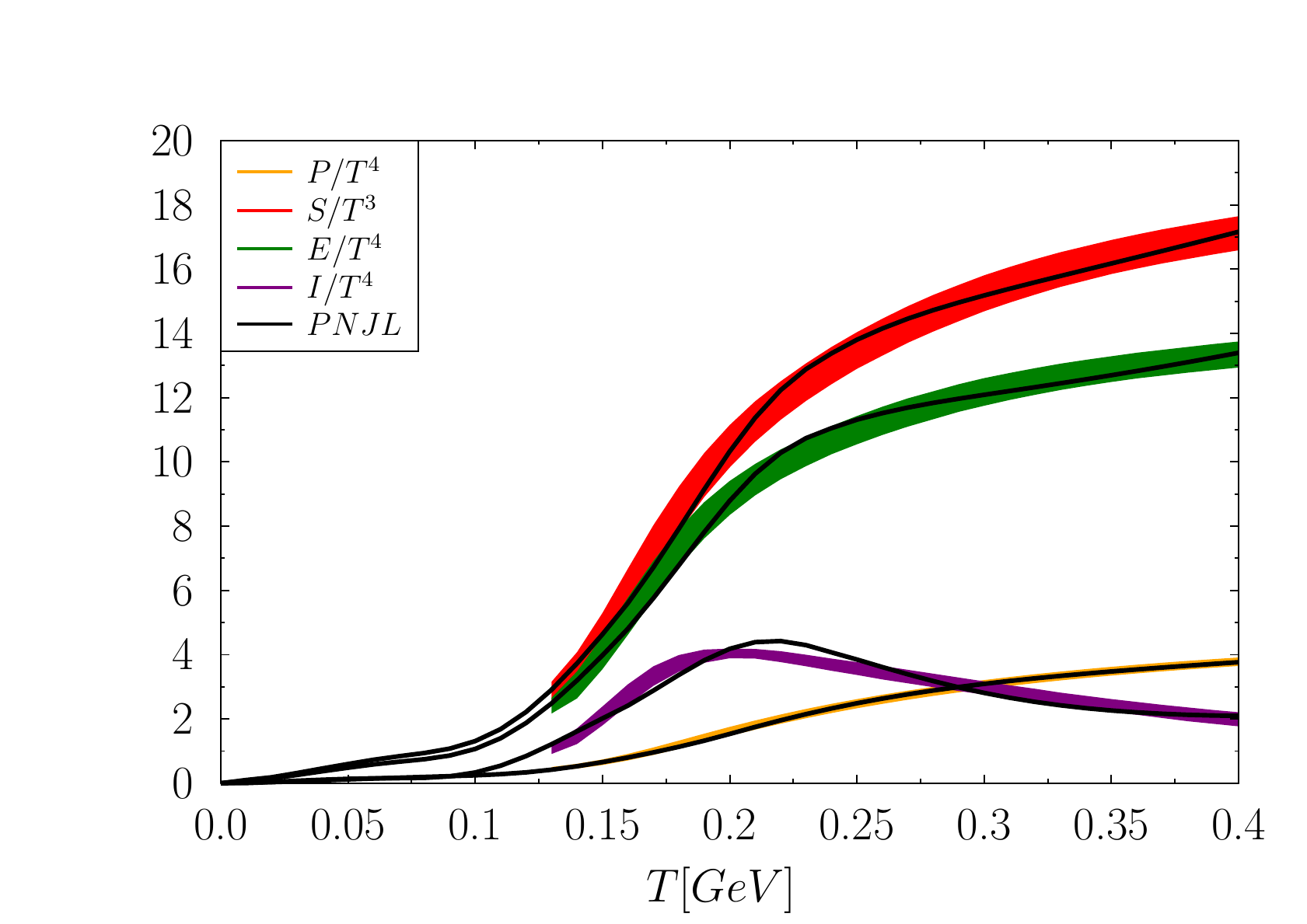}}
	    \caption{Pressure, entropy density, energy density and interaction measure calculated within the PNJL using the sum of Eqs \ref{OmegaMF} and \ref{potential meson}
for $\mu_q = 0$
(colored lines described in the legend) in comparison
 to the lattice QCD results \cite{Bazavov:2014pvz}, indicated as colored bands.}
	    \label{lattvsme}
	    \end{figure}
	    \end{centering}

\begin{centering}
\begin{figure}[h!]
\includegraphics[scale=0.5]{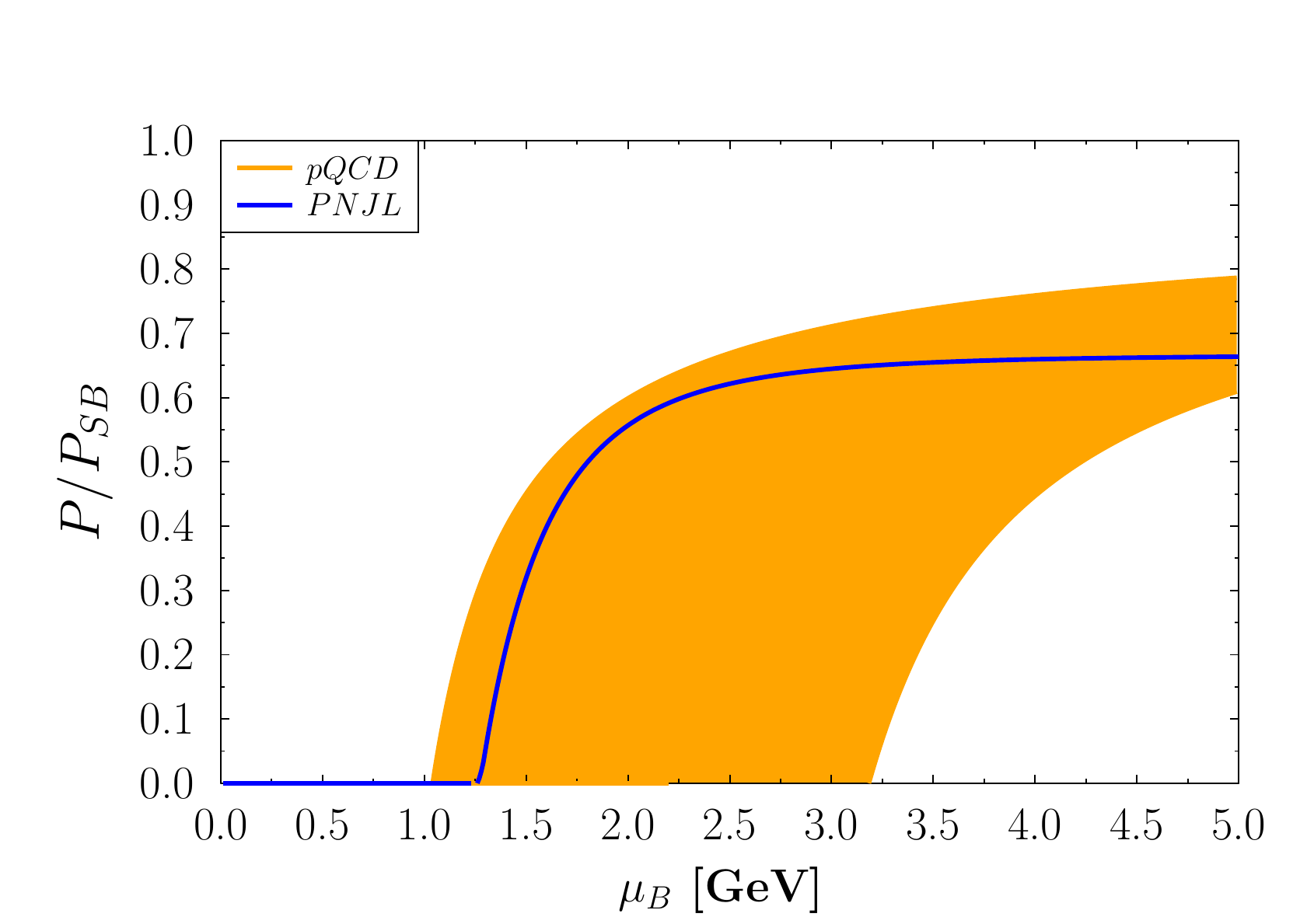}
\caption{The quark pressure divided by the pressure in the Stefan-Boltzmann limit as a function of $\mu_B$ for a temperature of $T = 0.001$ GeV. We compare pQCD calculations \cite{Kurkela:2016was} (orange area) with the result of our PNJL approach (blue line).}
\label{www}
\end{figure}
\end{centering}

The equation-of-state is needed to evaluate the entropy density, necessary to determine the shear viscosity to entropy ratio.  
Here we present the equation-of-state  of the improved PNJL model advanced  in \cite{Fuseau:2019zld} which  matches the lattice results of Ref.~\cite{Bazavov:2014pvz}.  
 The improved PNJL model differs from the standard PNJL model in two aspects:
	    \begin{itemize}
	    	\item  The grand potential includes next to leading order contributions in $N_c$ and contains therefore contributions from  mesons.
	    	\item A temperature dependent rescaling of the $T_0$ parameters of the standard Polyakov effective potential, see Eq. \ref{OmegaMF} and the Table \ref{table PNJL},  in order to phenomenologically reproduce the quark gluon interactions in the medium.
	    \end{itemize}
 In next to leading order the grand potential contains an additional term, $\Omega_M$, caused by the diagrams of the  order ${\cal O}(N_c=1)$ in the  $N_c$ expansion. 
 This term is given by \cite{Fuseau:2019zld,Torres-Rincon:2017zbr}:
    		\begin{align}
			&\Omega_M = -\frac{g_M}{8\pi^3}\int dpp^2\int\frac{ds}{\sqrt{s + \vec{k}^2}}\times \nonumber\\
			&\left[\frac{1}{\exp((\sqrt{s+\vec{k}^2}-\mu)/T-1)}\right.\nonumber\\
			&\left.+\frac{1}{\exp((\sqrt{s+\vec{k}^2}+\mu)/T-1)}\right]
			\delta(\sqrt{s},T,\mu_M).
			\label{potential meson}
		\end{align}
		Here $g_M$ is the degeneracy of the meson and $\delta(\sqrt{s},T,\mu_M)$ is the phase shift defined by
		\begin{equation}
	\delta(\sqrt{s},T,\mu_M)= -\text{Arg}[1-2K\Pi(\omega-\mu_M+i\epsilon,\vec{k})],
			\label{phase shift}
		\end{equation}
		where $\Pi$ is the polarization function of the meson. $\Omega_M$ represents a mesonic pressure that dominates at low temperature and is non negligible around $T_C$, as seen Fig. \ref{PQ}. 
As Fig. \ref{lattvsme} shows,   this equation-of-state reproduces the lattice results of Ref.~\cite{Bazavov:2014pvz} for vanishing chemical potential. Also
for  large chemical potentials and low temperatures, a region which is accessible for pQCD calculations,  we reproduce the pQCD calculations  of~\cite{Kurkela:2016was}
as shown in Fig.  \ref{www}.

	\subsection{Quark-quark scattering in the PNJL}
	
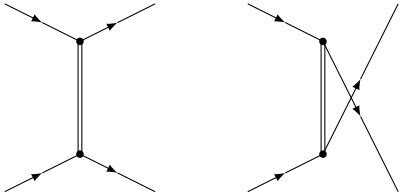
\begin{figure}[h!]
	    \begin{tabular}{cc}
		    \begin{tikzpicture}[scale=.5]
	    	\draw[->,>=latex] (0,0)--(1,-0.5);
	    	\draw(1,-0.5)--(2,-1);
	    	\draw[->,>=latex] (2,-1)--(3,-0.5);
	    	\draw(3,-0.5)--(4,0);
	    	
	    	\fill (2,-1) circle(0.1);
	    	
	    	\draw(1.95,-1)--(1.95,-4);
	    	\draw(2.05,-1)--(2.05,-4);
	    	
	    	\fill (2,-4) circle(0.1);
	    	
	    	\draw[->,>=latex] (0,-5)--(1,-4.5);
	    	\draw(1,-4.5)--(2,-4);
	    	\draw[->,>=latex] (2,-4)--(3,-4.5);
	    	\draw(3,-4.5)--(4,-5);
	    \end{tikzpicture} \hspace{1cm}  \begin{tikzpicture}[scale=.5]
	    	\draw[->,>=latex] (0,0)--(1,-0.5);
	    	\draw(1,-0.5)--(2,-1);
	    	\draw[->,>=latex] (2,-1)--(3,-3);
	    	\draw(3,-3)--(4,-5);
	    	
	    	\fill (2,-1) circle(0.1);
	    	
	    	\draw(1.95,-1)--(1.95,-4);
	    	\draw(2.05,-1)--(2.05,-4);
	    	
	    	\fill (2,-4) circle(0.1);
	    	
	    	\draw[->,>=latex] (0,-5)--(1,-4.5);
	    	\draw(1,-4.5)--(2,-4);
	    	\draw[->,>=latex] (2,-4)--(3,-2);
	    	\draw(3,-2)--(4,0);
	    \end{tikzpicture}
	    \end{tabular}
\caption{ The Feynman diagrams for the $t-$ and $u-$ channels which contribute 
to the quark-quark cross sections.}
	    \label{FDx}
	    \end{figure}

	There are two possible Feynman diagrams for quark-quark scattering, 
	the $t-$ and $u-$ channels as indicated in Fig. \ref{FDx}.	
The associated  squared of the matrix elements for the $t-$ and $u-$ channels 
and their interference term are defined as 
	    \begin{equation}
	    \frac{1}{4N_C^2}\sum_{s,c}|M_u|^2 = |\mathscr{D}_u^S|^2u^+_{14}u^+_{23} +|\mathscr{D}_u^P|^2u^-_{14}u^-_{23},
	    \end{equation}
	    \begin{align}
	    \frac{1}{4N_C^2}\sum_{s,c}&|M_{ut}| = \frac{1}{4N_C}[\nonumber\\
	    &\mathscr{D}_t^S\mathscr{D}_u^{S*}( t^+_{13}t^+_{24} - s^+_{12}s^+_{34} + u^+_{14}u^+_{23} )\nonumber\\ &- \mathscr{D}_t^S\mathscr{D}_u^{P*}( t^+_{13}t^+_{24} - s^-_{12}s^-_{34} + u^-_{14}u^-_{23} ) \nonumber\\ &- \mathscr{D}_t^P\mathscr{D}_u^{S*}( t^-_{13}t^-_{24} - s^-_{12}s^-_{34} + u^+_{14}u^+_{23} )\nonumber\\ &+ \mathscr{D}_t^P\mathscr{D}_u^{P*}( t^-_{13}t^-_{24} - s^+_{12}s^+_{34} + u^-_{14}u^-_{23} )],
	    \end{align}
   \begin{equation}
	    \frac{1}{4N_C^2}\sum_{s,c}|M_t|^2 = |\mathscr{D}_t^S|^2t^+_{13}t^+_{24} +|\mathscr{D}_t^P|^2t^-_{13}t^-_{24}.
	    \end{equation}
Here	
	\begin{equation}
	t^\pm_{ij} = t - (m_i\pm m_j)^2,  
	\end{equation}
	\begin{equation}
	u^\pm_{ij} = u - (m_i\pm m_j)^2,
	\end{equation}	
	\begin{equation}
	s^\pm_{ij} = s - (m_i\pm m_j)^2 .
	\end{equation}
$\mathscr{D}^S$ and $\mathscr{D}^P$ are the propagators of the exchanged scalar and pseudoscalar meson, respectively, see Eq.\ref{amplimesonex}. 

Table \ref{table qq} lists 
the mesons which can be exchanged in the $t-$ and $u-$ channels in the different 
quark-quark cross sections \cite{Rehberg:1996vd}.

    	\begin{table}[h!]
	    \centerline{\begin{tabular}{|c|c|c|}
	    \hline
	    Process & Exchanged mesons  & Exchanged mesons \\
	        & in $u$-channel &  in $t$-channel \\
	    \hline
	    $ud\rightarrow ud$ & $\pi$, $\sigma_\pi$ & $\pi$, $\eta$, $\eta'$, $\sigma_\pi$, $\sigma$, $\sigma'$\\
	    \hline
	    $uu\rightarrow uu$ & $\pi$, $\eta$, $\eta'$, $\sigma_\pi$, $\sigma$, $\sigma'$ & $\pi$, $\eta$, $\eta'$, $\sigma_\pi$, $\sigma$, $\sigma'$\\
	    \hline
	    $us\rightarrow us$ & K, $\sigma_K$ & $\eta$, $\eta'$, $\sigma$, $\sigma'$\\
	    \hline
	    $ss\rightarrow ss$ & $\eta$, $\eta'$, $\sigma$, $\sigma'$ & $\eta$, $\eta'$, $\sigma$, $\sigma'$\\
	    \hline	    
	    \end{tabular}}
	    \caption{Mesons which can be exchanged in the $t-$ and $u-$ channels 
	    in the different quark-quark cross sections.}
	    \label{table qq}
	    \end{table}

\subsection{Quark-antiquark scattering in the PNJL}

\begin{figure}[h!]
	    \begin{tabular}{cc}
		    \begin{tikzpicture}[scale=.5]
	    	\draw[->,>=latex] (0,0)--(1,-0.5);
	    	\draw(1,-0.5)--(2,-1);
	    	\draw[->,>=latex] (2,-1)--(3,-0.5);
	    	\draw(3,-0.5)--(4,0);
	    	
	    	\fill (2,-1) circle(0.1);
	    	
	    	\draw(1.95,-1)--(1.95,-4);
	    	\draw(2.05,-1)--(2.05,-4);
	    	
	    	\fill (2,-4) circle(0.1);
	    	
	    	\draw[-<,>=latex] (0,-5)--(1,-4.5);
	    	\draw(1,-4.5)--(2,-4);
	    	\draw[-<,>=latex] (2,-4)--(3.,-4.5);
	    	\draw(3,-4.5)--(4,-5);
	    \end{tikzpicture} \hspace{1cm}\begin{tikzpicture} [scale=.5]
	    	\draw[->,>=latex] (0,0)--(1,-1);
	    	\draw(1,-1)--(2,-2);
	    	\draw[-<,>=latex] (0,-4)--(1,-3);
	    	\draw(1,-3)--(2,-2);

	    	\fill (2,-2) circle(0.1);
	    	
	    	\draw(2,-2.05)--(4,-2.05);
	    	\draw(2,-1.95)--(4,-1.95);
	    	
	    	\fill (4,-2) circle(0.1);
	    	
	    	\draw[->,>=latex] (4,-2)--(5,-1);
	    	\draw(5,-1)--(6,0);
	    	\draw[-<,>=latex] (4,-2)--(5,-3);
	    	\draw(5,-3)--(6,-4);
	    \end{tikzpicture}
	    \end{tabular}
	    \caption{The Feynman diagrams for the $t-$ and $s-$ channels 
	    which contribute to the quark-antiquark cross sections.}
	    \label{FDst}
	    \end{figure}
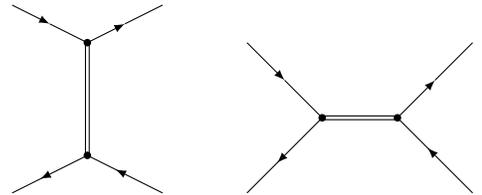

 For quark-antiquark scattering only $s-$ and $t-$ channels are possible.
The corresponding diagrams are shown in Fig. \ref{FDst}.
	    
The corresponding squared of the matrix elements for the $s-$ and $t-$ channels 
and their interference term are given by
	    \begin{equation}
	    \frac{1}{4N_C^2}\sum_{s,c}|M_s|^2 = |\mathscr{D}_s^S|^2s^+_{12}s^+_{34} +|\mathscr{D}_s^P|^2s^-_{12}s^-_{34},
	    \end{equation}
	   \begin{align}
	    \frac{1}{4N_C^2}\sum_{s,c}&|M_{st}| = \frac{1}{4N_C}[\nonumber\\
	    &\mathscr{D}_s^S\mathscr{D}_t^{S*}( s^+_{12}s^+_{34} - u^+_{14}u^+_{23} + t^+_{13}t^+_{24} )\nonumber\\ &- \mathscr{D}_s^S\mathscr{D}_t^{P*}( s^+_{12}s^+_{34} - u^-_{14}u^-_{24} + t^-_{13}t^-_{24} ) \nonumber\\ &- \mathscr{D}_s^P\mathscr{D}_t^{S*}( s^-_{12}s^-_{34} - u^-_{14}u^-_{23} + t^+_{13}t^+_{24} )\nonumber\\ &+ \mathscr{D}_s^P\mathscr{D}_t^{P*}( s^-_{12}s^-_{34} - u^+_{14}u^+_{23} + t^-_{13}t^-_{24} )],
	    \end{align}
	    \begin{equation}
	    \frac{1}{4N_C^2}\sum_{s,c}|M_t|^2 = |\mathscr{D}_t^S|^2t^+_{13}t^+_{24} +|\mathscr{D}_t^P|^2t^-_{13}t^-_{24}.
	    \end{equation}
Table \ref{process qqb} presents the mesons which can be exchanged 
in the $s-$ and $t-$ channels in the different quark-(anti-)quark cross sections 
\cite{Rehberg:1996vd}.
	\begin{table}[h!]
	    \centerline{\begin{tabular}{|c|c|c|}
	    \hline
	    Process & Exchanged mesons  & Exchanged mesons \\
	        & in s-channel &  in t-channel \\
	    \hline
	    $u\bar{d}\rightarrow u\bar{d}$ & $\pi$, $\sigma_\pi$ & $\pi$, $\eta$, $\eta'$, $\sigma_\pi$, $\sigma$, $\sigma'$\\
	    \hline
	    $u\bar{u}\rightarrow u\bar{u}$ & $\pi$, $\eta$, $\eta'$, $\sigma_\pi$, $\sigma$, $\sigma'$ & $\pi$, $\eta$, $\eta'$, $\sigma_\pi$, $\sigma$, $\sigma'$\\
	    \hline
	    $u\bar{u}\rightarrow d\bar{d}$ & $\pi$, $\eta$, $\eta'$, $\sigma_\pi$, $\sigma$, $\sigma'$ & $\pi$, $\sigma_\pi$\\
	    \hline
	    $u\bar{s}\rightarrow u\bar{s}$ & K, $\sigma_K$ & $\eta$, $\eta'$, $\sigma$, $\sigma'$\\
	    \hline
	    $u\bar{u}\rightarrow s\bar{s}$ & $\eta$, $\eta'$, $\sigma$, $\sigma'$ & K, $\sigma_K$\\
	    \hline
	    $s\bar{s}\rightarrow u\bar{u}$ & $\eta$, $\eta'$, $\sigma$, $\sigma'$ & K, $\sigma_K$\\
	    \hline
	    $s\bar{s}\rightarrow s\bar{s}$ & $\eta$, $\eta'$, $\sigma$, $\sigma'$ & $\eta$, $\eta'$, $\sigma$, $\sigma'$\\
	    \hline	    
	    \end{tabular}}
	    \caption{Mesons which can be exchanged in the $s-$ and $t-$ channels in the different quark-antiquark cross sections.}
	     \label{process qqb}
	     \end{table}

\subsection{Integration boundaries and the total cross sections}
	
The differential cross sections for the quark-quark and quark-antiquark scattering
for $t-$, $u-$ channels and $t-$, $s-$ channels, respectively, and are given by the following expressions:	
	\begin{equation}
	\frac{d\sigma}{dt} = \frac{1}{16\pi s^+_{12}s^-_{12}}\frac{1}{4N_C^2}\sum_{s,c}|\mathscr{M}_{s/u} - \mathscr{M}_t|^2.
	\end{equation}	
The total cross section in a thermal medium is obtained by integration over $t$:
		\begin{equation}
	\sigma = \int_{t_-}^{t^+} dt \frac{d\sigma}{dt}[1-f_F(E_3,T,\mu)][1-f_F(E_4,T,\mu)],
	\end{equation}
where $1-f_F$ is the Pauli blocking factor for the fermions 
due to the fact that some of the final states are already occupied by 
other quarks(antiquarks).	
The limits of the integrations are defined as
	\begin{align}
	t_\pm &= m_1^2 + m_3^2 - \frac{1}{2s}(s+ m_1^2 - m_2^2)(s + m_3^2 - m_4^2)\nonumber\\ &\pm 2 \sqrt{\frac{(s + m_1^2 - m_2^2)^2}{4s}-m_1^2}	\sqrt{\frac{(s + m_3^2 - m_4^2)^2}{4s}-m_3^2}.
	\end{align}
where $m_1$ and $m_2$ are the masses of the particles in the entrance channel and $m_3$ and $m_4$ of those in the exit channel.

	\subsection{Results for elastic cross section}
	
	 \begin{figure}[h!]
	 \centering
	 \includegraphics[scale=0.5]{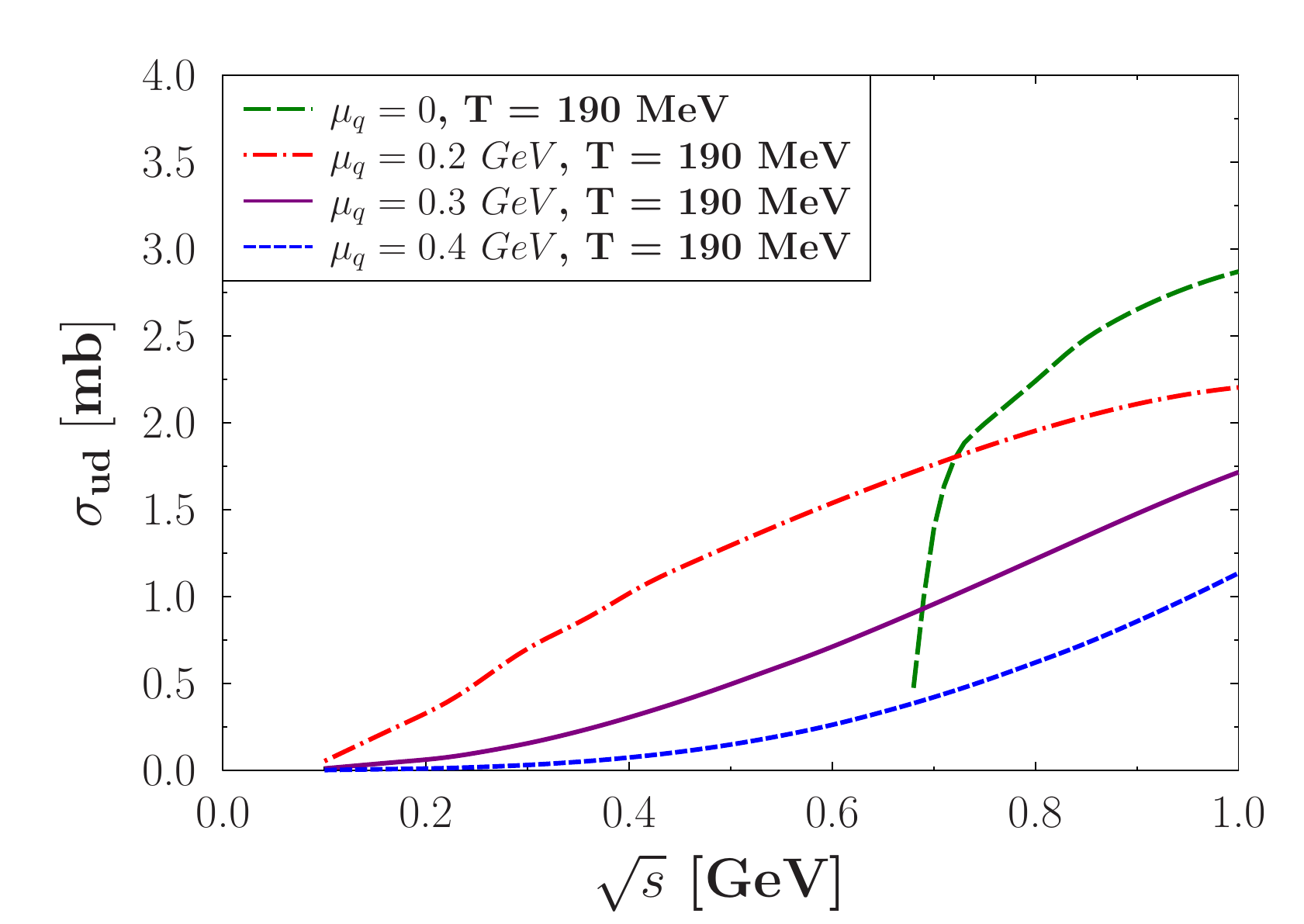}\ \includegraphics[scale=0.5]{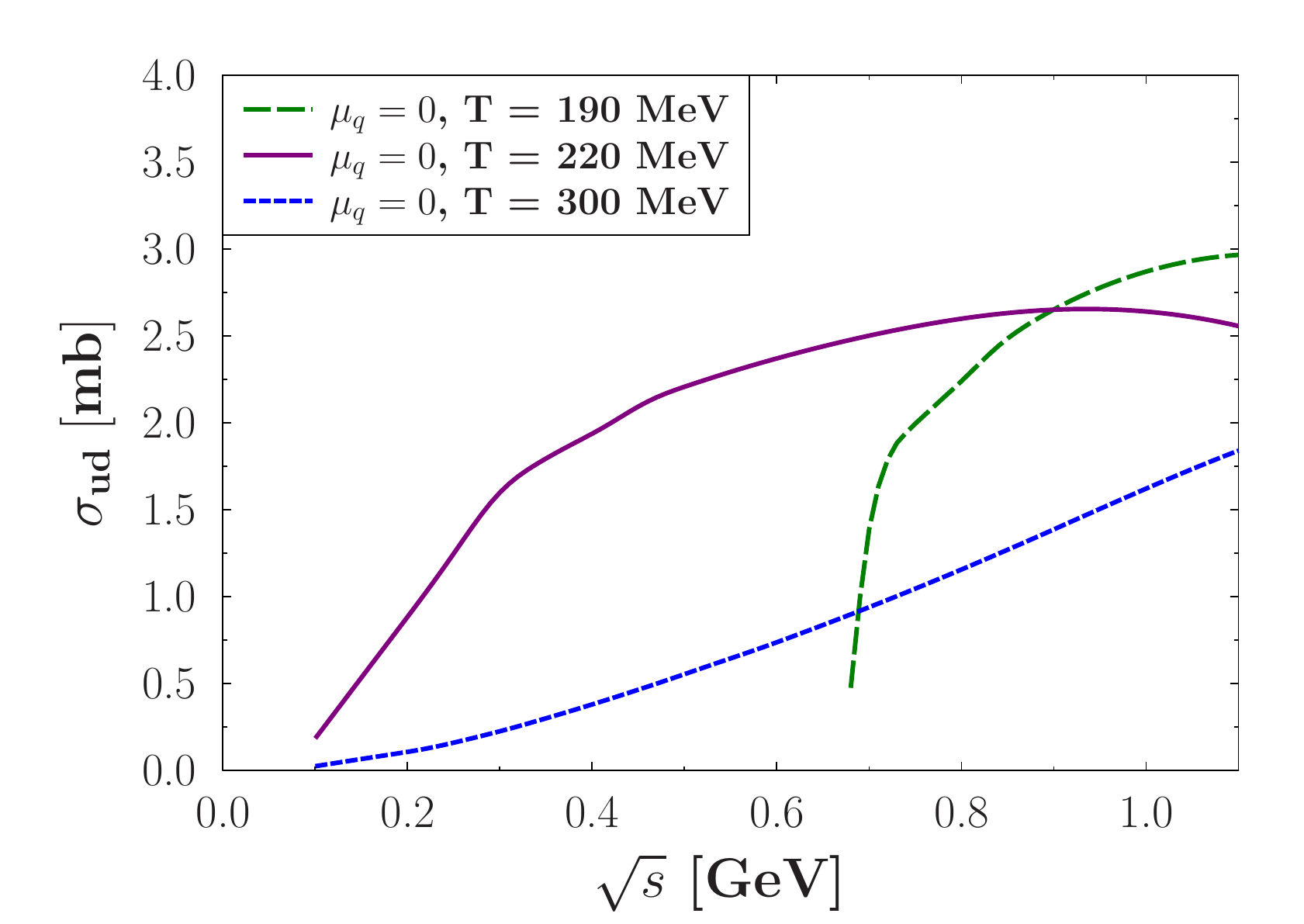}
	 \caption{The cross section $\sigma_{ud}$  versus $\sqrt{s}$ 
	 at $T=190$ MeV  for  $\mu_q = 0,\ 0.1,\ 0.2$ GeV (upper) 
	 and at $\mu_q=0$ for $T=$190, 220 and 300  MeV (lower).}
	 \label{cs qq}
	 \end{figure}
In Fig. \ref{cs qq} we present the quark-quark cross sections for $ud\to ud$
elastic scattering versus $\sqrt{s}$ at $T=190$ MeV  
for different $\mu_q = 0,\ 0.1,\ 0.2$ GeV (upper plot) and at $\mu_q=0$ 
for different $T=$190, 220 and 300  MeV (lower plot).
As follows from Fig. \ref{cs qq} these cross sections are rather small and show 
a smooth behavior versus the center of mass energy. They are decreasing 
with increasing temperature and chemical potential what is expected 
because the mass of the exchanged meson and its decay width increase with temperature. 
At low temperature and chemical potential the masses of the quarks are large  
and tend to vanish at large $T$ and $\mu_q$. Consequently, the threshold,  
given by  $\sqrt{s}_{thr} = Max(m_{in}^a+m_{in}^b$, $m_{out}^a+m_{out}^b)$,  is high at low $\mu_q$ and low $T$, as one can see from Fig. \ref{cs qq} as well. 
	 
The more interesting processes are quark-antiquark collisions. In this case the 
$s-$ channel allows for a resonance of the exchanged meson with the incoming quarks 
which leads to a large peak in the cross sections.
	 
	 \begin{figure}[h!]
	 \centering
	 \includegraphics[scale=0.5]{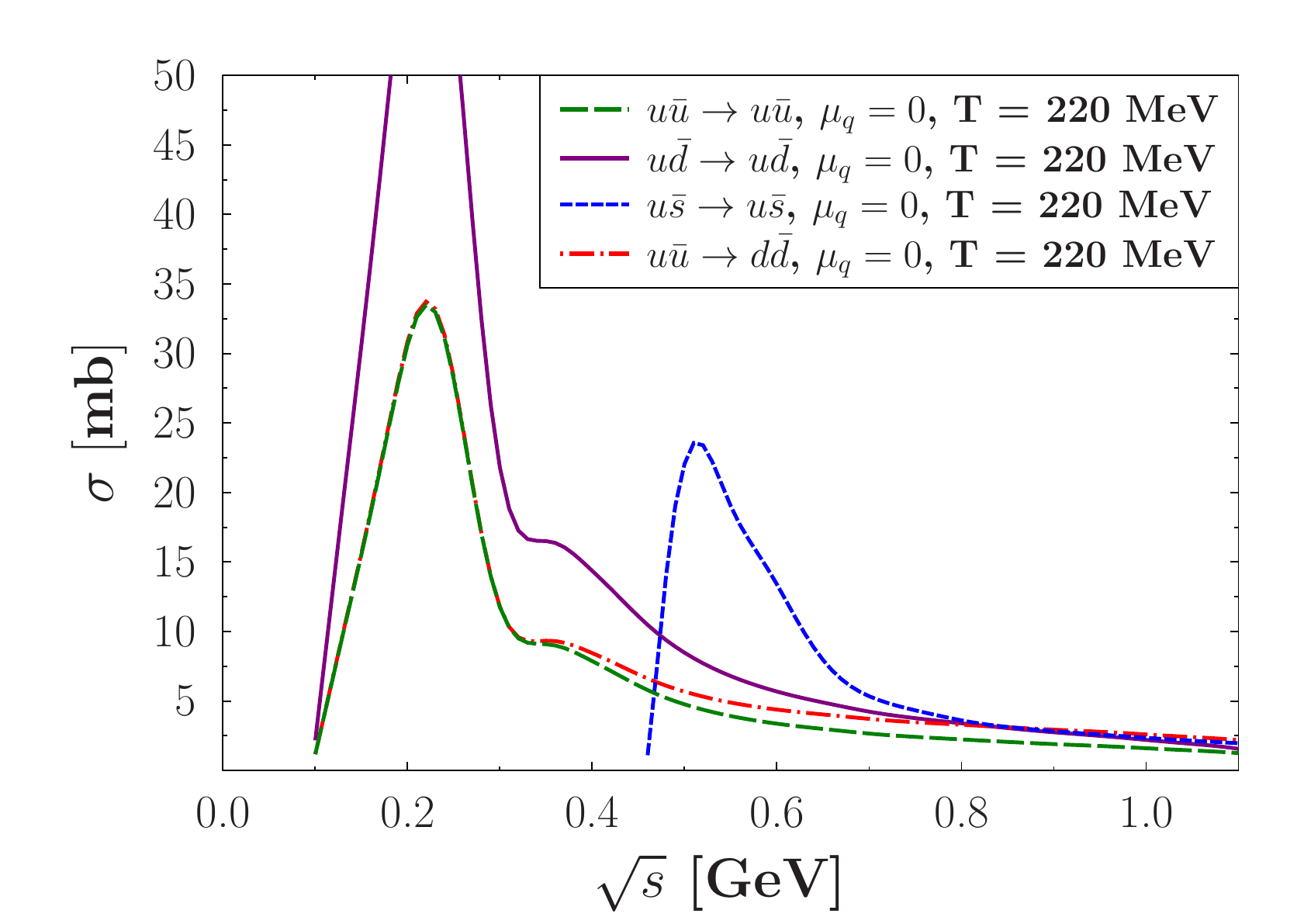}\ \includegraphics[scale=0.5]{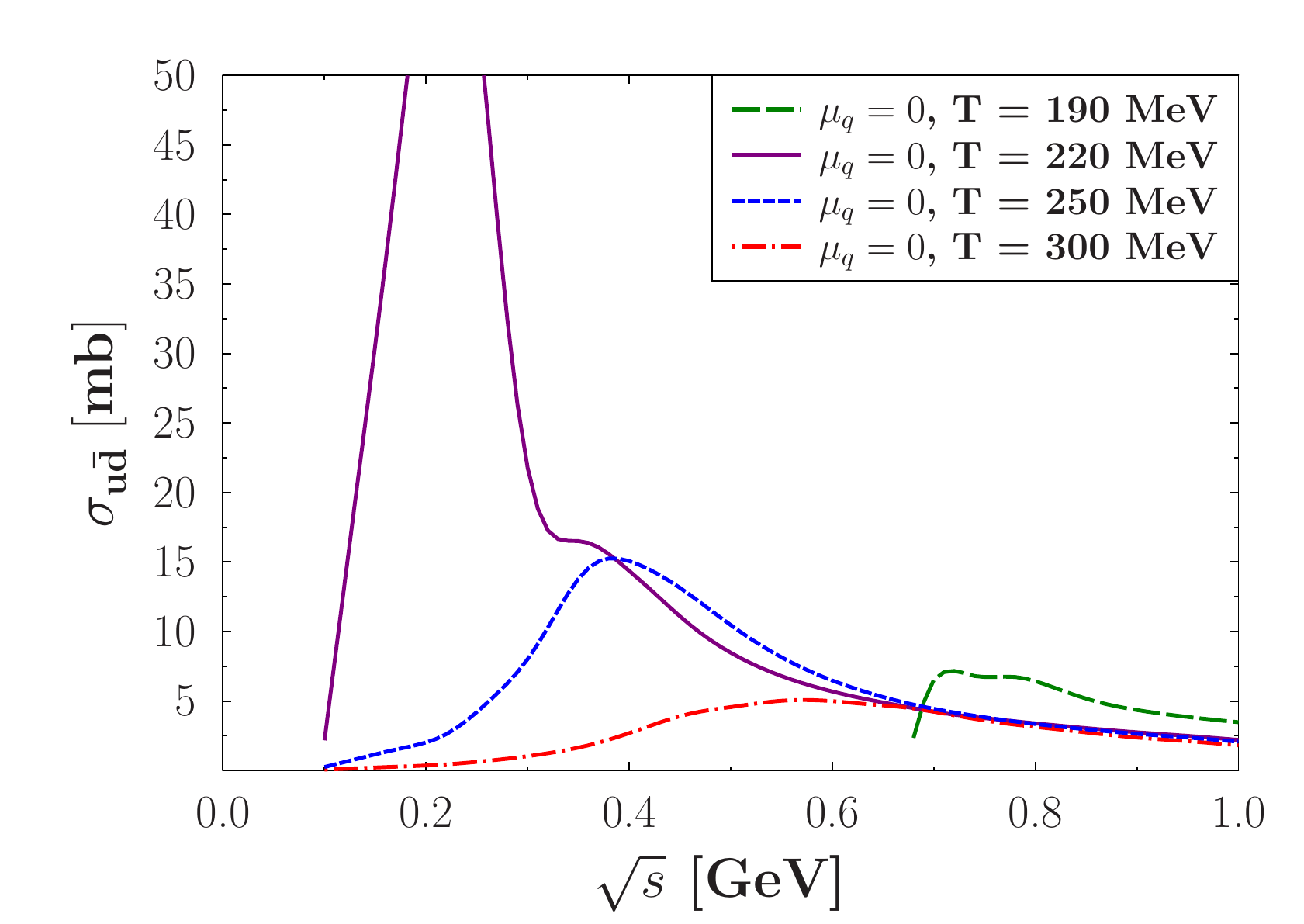}
	 \caption{ The resonance behaviour of the $u\bar{u}\rightarrow u\bar{u}$, $u\bar{d}\rightarrow u\bar{d}$, $u\bar{u}\rightarrow d\bar{d}$ and $u\bar{s}\rightarrow u\bar{s}$ cross sections versus $\sqrt{s}$ at  $T = $ 220 GeV and $\mu=0$ (upper) and 
at $\mu_q=0$ for $T=190$, 220, 250 and 300 MeV.}
	 \label{udb}
	 \end{figure}

Fig.~\ref{udb} (upper part) displays the cross section at different channels showing a resonance behaviour. The $u\bar{s}\rightarrow u\bar{s}$ resonance is lower than the others because the strange quark is heavier than the u and d quarks at $\mu_q=0$ and $T=200$ MeV.
The other resonances differ only by their flavour factors \cite{Blanquier:2013eja}. 
The $u\bar{d}\rightarrow u\bar{d}$ channel has the largest factor, 
$u\bar{u}\rightarrow u\bar{u}$ has a lower factor than $u\bar{u}\rightarrow d\bar{d}$ 
but allows for a $\eta$ meson exchange which is not the case for the 
$u\bar{u}\rightarrow d\bar{d}$ channel.

The behavior of the $u\bar{d}\rightarrow u\bar{d}$ cross section for different temperatures is displayed in Fig.~\ref{udb} (lower part). 
One can see that the resonance is shifted to the right when the temperature increases. 
Since the mass of the mesons increases with temperature, the cross section 
with the pion in the $s-$ channel becomes resonant at the corresponding $\sqrt{s}$. 
The peak becomes lower with increasing temperature and disappears finally at large temperatures since the decay width of the pion becomes larger  with increasing temperature, see Fig.~\ref{mes mas cs}.
The kinematic threshold forbids any resonance state below the Mott temperature. 
This explains the flatness of the $u\bar{d}\rightarrow u\bar{d}$ cross section at 
$T=190$ MeV.

		 \begin{figure}[h!]
	 \centering
 \includegraphics[scale=0.5]{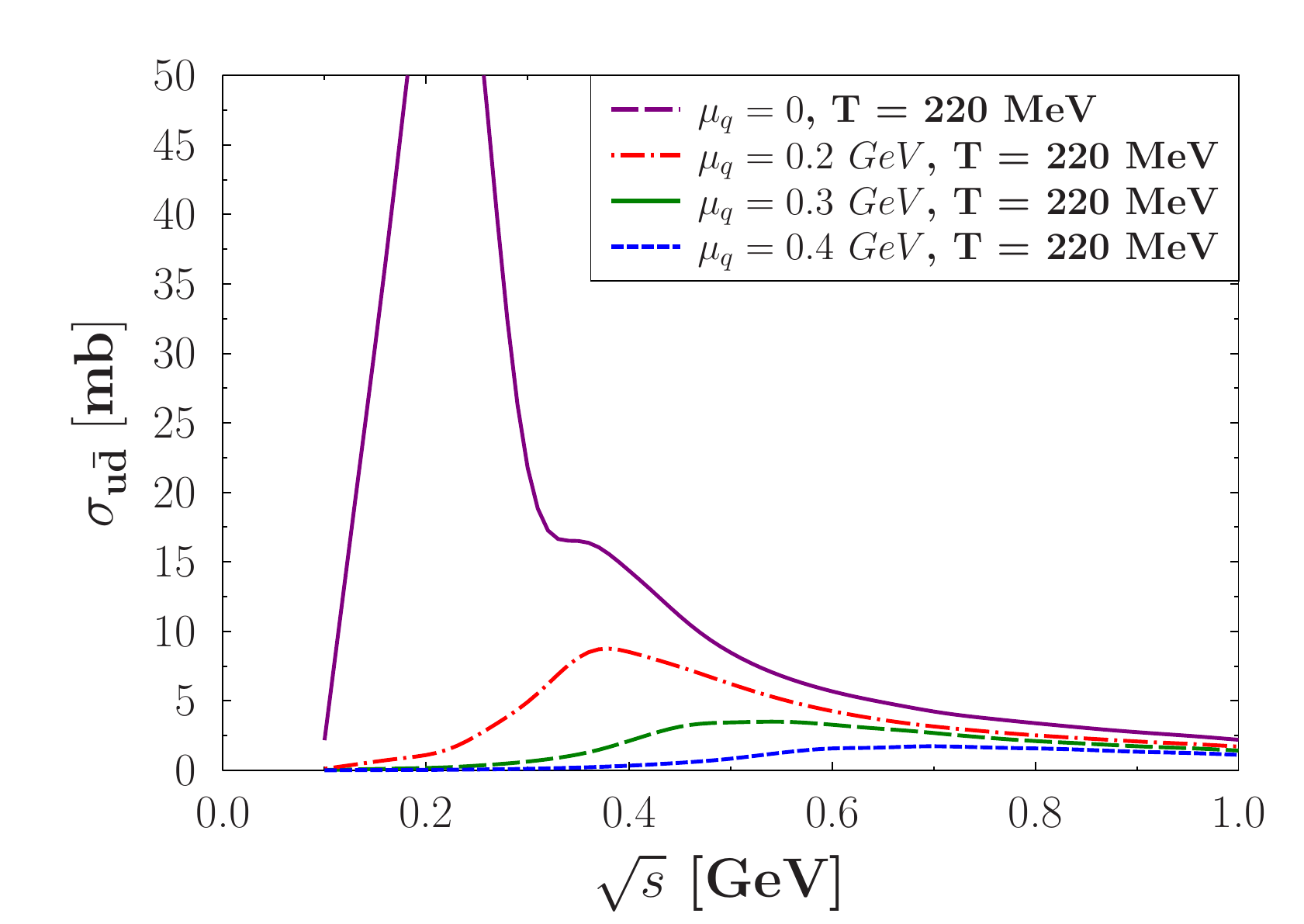}\
  \includegraphics[scale=0.5]{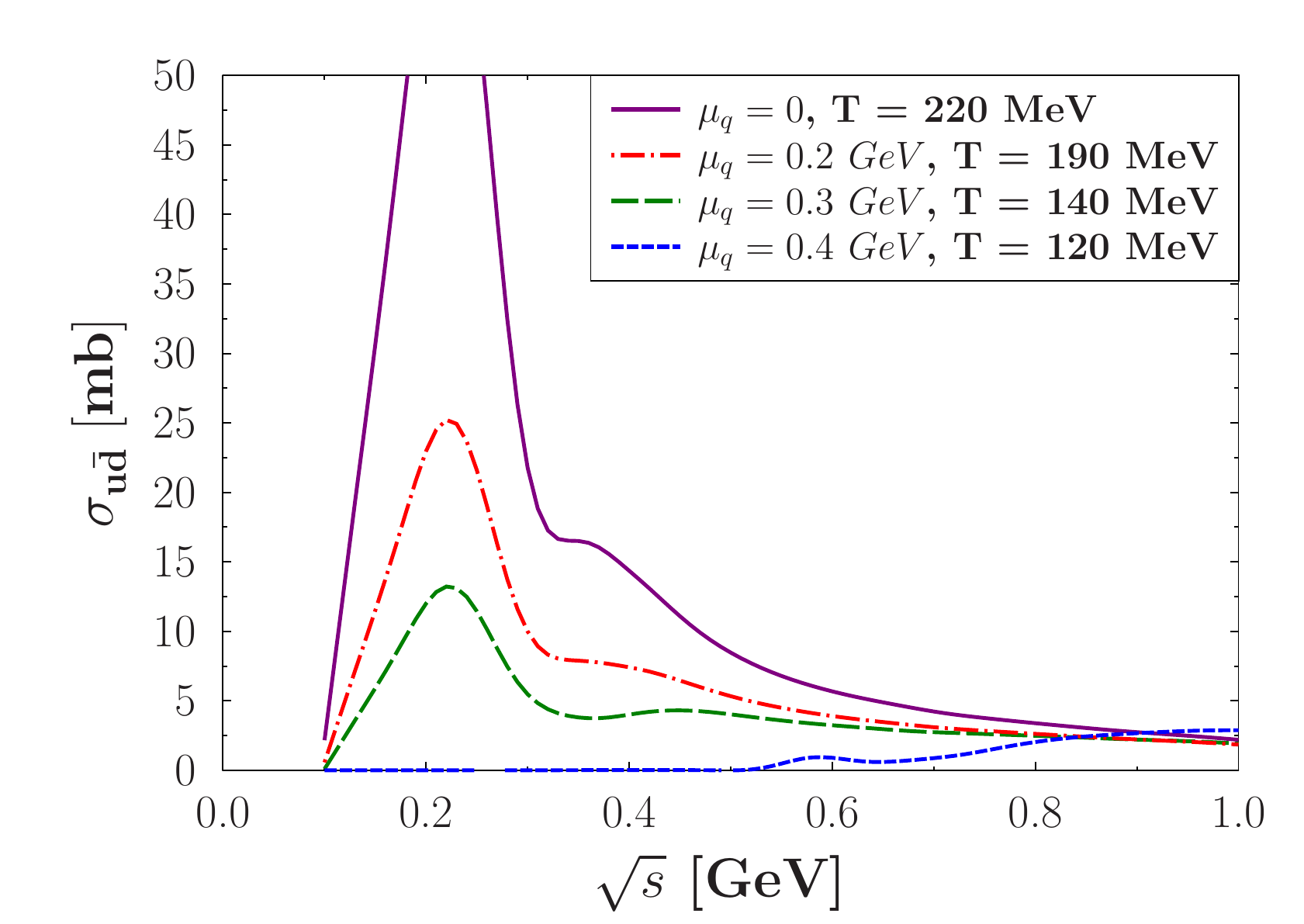}
	 \caption{The resonant $u\bar{d}$ cross section versus $\sqrt{s}$
	 at $T=$220 MeV  for $\mu_q=$0, 200, 300 and 400 MeV (upper) 
	 and for different combinations of $T, \mu_q$ (lower).}
	 \label{udbmu}
	 \end{figure} 

Fig.~\ref{udbmu} shows the behavior of the resonance peak with increasing chemical potential. For a given temperature, the mass of the pion becomes larger with increasing chemical potential and the peak is shifted to a smaller value of the temperature.
Beyond the critical endpoint $\mu_{CEP} = 0.32$ GeV, the cross section is flat and no resonance behaviour shows up anymore. \\

\begin{figure}[h!]
	 \centering
 \includegraphics[scale=0.5]{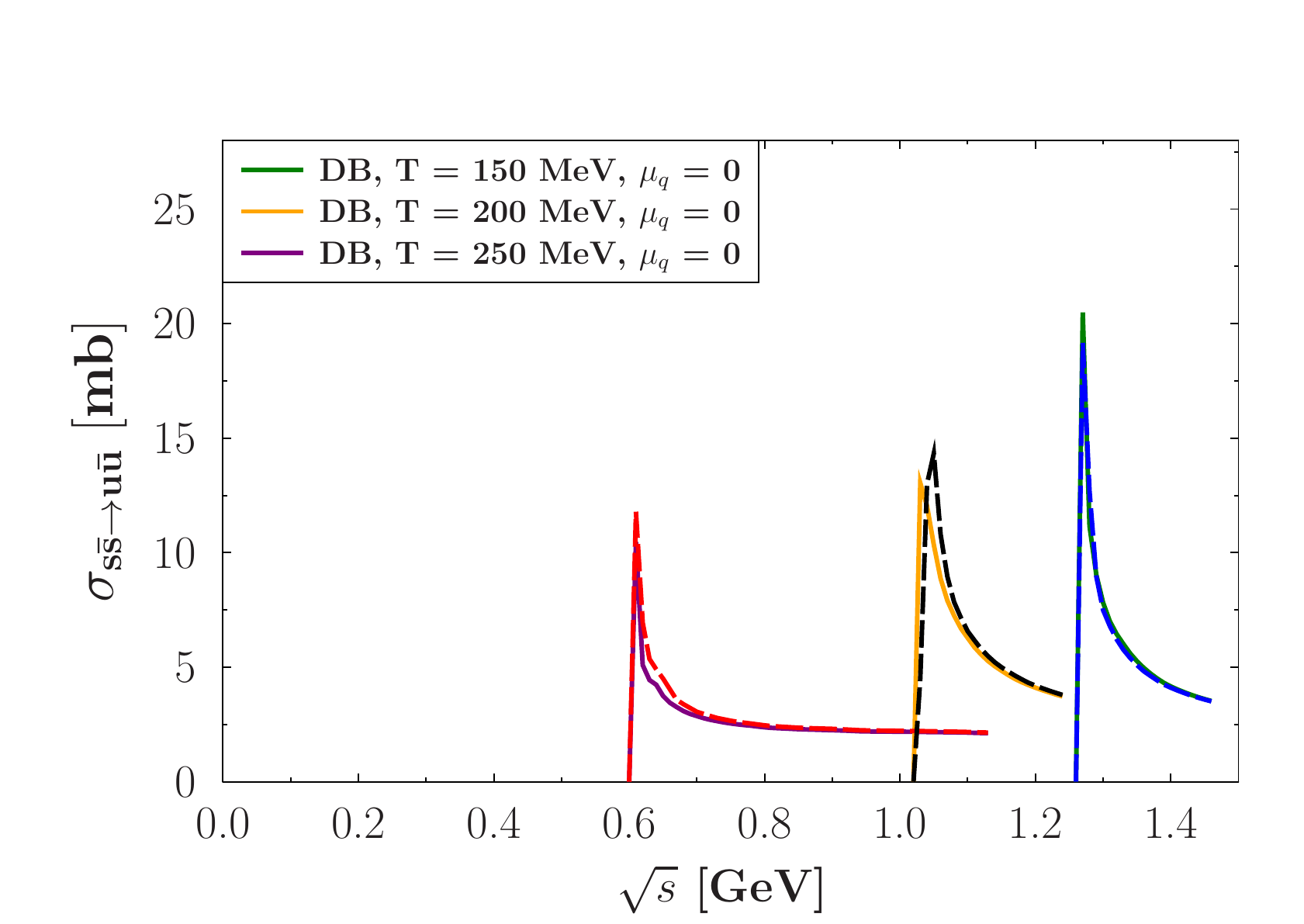}
	 \caption{The cross section for the $s\bar{s}\rightarrow u\bar{u}$ channel 
	 calculated by detailed balance (solid lines, DB) for $T =$ 150, 200, 250 MeV at 
	 $\mu_q$ = 0  as compared to the direct  numerical calculation (dashed lines).}
	 \label{detbal}
	 \end{figure}

The calculation of  the two cross sections $s\bar{s}\rightarrow u\bar{u}$ and $u\bar{u}\rightarrow s\bar{s}$ can be double checked since they obey detailed balance:
\begin{equation}
\sigma_{cd\rightarrow ij} (s)=\frac{p^{2\ cm}_{ij}(s)}{p^{2\ cm}_{cd}(s)} \sigma_{ij\rightarrow cd} (s).
\label{DB}
\end{equation}
Fig.~\ref{detbal} shows the cross section for the $s\bar{s}\rightarrow u\bar{u}$
channel at $\mu_q=0$ and $T =$ 150, 200, 250 MeV calculated directly (dashed lines)
and by detailed balance (\ref{DB})(solid lines). One can see that both calculations show 
a good agreement with each other.

\section{\label{sec3}Transport coefficients in the PNJL model}
	    
\subsection{Transport coefficients in the relaxation time approximation}

	   In the relativistic kinetic theory one can determine the transport coefficients with help of  the relaxation time approximation of the Boltzmann equation for the quasiparticles with dynamical masses $M_i(T,\mu_q)$ \cite{Romatschke:2011qp,Chakraborty11,Alqahtani:2015qja,Kapusta}: 
	   	\begin{align}
	k_i^{\mu }\partial_\mu f_{i} + \frac{1}{2}\partial^\mu M_i^2 \partial_{(k_{i,\, \mu})} f_{i} = \sum\limits_{j \, = \, 1}^{N_{\text{species}}}C_{ij}(x,k), \label{eq:BoltzmannEq}
	\end{align}
 where $C_{ij}(x,k)$ is the 2-body collision term which contains only quasi-elastic $2\leftrightarrow2$ scatterings, while the second term contains 
 $F^\mu_i=\partial^\mu M_i$ and is an external force attributed to the 
 residual mean field interaction due to the medium dependent effective masses 
 $M_i(T,\mu_q)$. 
 
	   	In order to evaluate transport coefficients we consider a small departure from equilibrium, where the distribution function can be expressed as 
\bea
f_i(x,k,t) = f^{(0)}_i(x,k,t)+f^{(1)}_{i}(x,k,t)= \nonumber\\ f^{(0)}_i(x,k,t)(1 + \delta f_{i}(x,k,t) ).
  \eea
  $f^{(0)}_i(x,k,t)$ is the local equilibrium distribution function, $f^{(1)}_{i}(x,k,t)$ contains $\delta f_{i}(x,k,t)$, which is  the non equilibrium part to first order in gradients.
  Quark systems in equilibrium can be described by the Fermi-Dirac distribution function: 
    	   	\begin{equation}
	f^{(0)}_i(E_i,T,\mu_q) = \frac{1}{e^{(E_i\pm \mu_q)/T}+1}, 
	   \end{equation}
	   where $E_i=\sqrt{p_i^2+m_i^2}$ is the  on-shell quark energy, $T$ is the temperature and $\mu_q=\mu_B/3$ for the light quarks, $\mu_s=0$ for the strange quark. 
	   The (anti-)quark density is defined as 
\bea
n_i(T,\mu_q)= d_q \int \frac{d^3p}{(2\pi)^3} f^{(0)}_i, 
\label{ndens}
\eea 
	   where $i=u,d,s,\bar u, \bar d, \bar s$ and $d_q=2 \times N_c$ is degeneracy factor for (anti-)quarks. 
	   
	   In order to take into account the Polyakov loop contributions we  use the modified Fermi-Dirac distribution:
	   \begin{align}
f^{\phi}_i = \frac{\phi e^{-(E_i\mp \mu)/T}+ 2 \overline{\phi} e^{-2(E_i\mp \mu)/T}+e^{-3(E_i\mp \mu)/T}}{1+3\phi e^{-(E_i\mp \mu)/T}+3\overline{\phi} e^{-2(E_i \mp \mu)/T}+e^{-3(E_i \mp \mu)/T}},
     \label{eq:fpol_qbar}
     \end{align}
     where $i=q,\overline{q}$. The minus sign refers to quarks ($i=q$), while the plus sign refers to antiquarks ($i=\overline{q}$). For antiquarks we have to exchange $\phi$ and $\bar \phi$.

     In the QGP phase the modified distributions approach the standard Fermi-Dirac distributions for $\phi \rightarrow 1$, while in the hadronic phase, for $\phi \rightarrow 0$, we get
a distributions with three times the quark energy in the exponent, which can be interpreted as a
Fermi-Dirac distribution function of a particle with three times the quark mass. 

	 In the relaxation time approximation to first order in the deviation from equilibrium the collision term is given by \cite{Anderson:1974rta}
	 	\begin{align}
		\sum\limits_{j \, = \, 1}^{N_{\text{species}}} \mathcal{C}_{ij}^{(1)}[f_{i}] = -\frac{E_{i}}{\tau_{i}} \left( f_{i} - f^{(0)}_{i}\right) = -\frac{E_{i}}{\tau_{i}} f^{(1)}_{i} + \mathcal{O}(\mathrm{Kn}^2) ,
	\end{align}
	 where $\tau_i$ is the relaxation time in the heat bath rest system for the particle species $'i'$, $\mathrm{Kn}\sim l_{micro}/L_{macro}$  is the Knudsen number which denotes  the  ratio  between the relevant microscopic/transport length scales. $l_{micro}$  is in our case the  mean free path $\lambda$, and the macroscopic scale $L_{macro}$ is the characteristic length of the system.

\subsection{Quark relaxation time}

 The RTA is often use in the framework of effective models for the estimation of transport coefficients in the QGP phase. It is worth to note that the results of
transport calculations depend not only on the EoS, which can be fitted to the lQCD results, but also (if no local equilibrium is assumed) on transport coefficients and therefore 
on the method of how to evaluate quark and gluon relaxation times. 

In this section we apply two different approaches for the calculation
of the quark relaxation time, which are commonly used in the literature:
1) the so called 'averaged transition rate' defined via the thermal averaged  
quark-quark and quark-antiquark PNJL cross sections
and 
2) the 'weighted' thermal averaged  quark-quark and quark-antiquark PNJL cross sections.
As will be demonstrated later, the differences between  both method are quite
essential and influence  substantially the final results for the transport coefficients.
 
\subsubsection{Method 1 for the quark relaxation time} 

We start with the estimation of the quark relaxation time through 
the averaged interaction rate, related to the thermal averaged  quark-quark and quark-antiquark PNJL cross sections, advanced in \cite{Hosoya:1983xm,Chakraborty11,Kapusta,Berrehrah:2014kba}. 
The momentum dependent relaxation time can be expressed through the on-shell 
interaction rate in the medium rest system where the incoming quark has a four-momentum $P_i=(E_i,\mathbf{p}_i)$ :
 \bea
 &\tau^{-1}_i&(p_i,T,\mu_q  )= \Gamma_{i}(p_i,T,\mu_q ) \label{Gamma_on}  \\
	&=& \frac{1}{2E_i} \sum_{j=q,\bar{q}} \frac{1}{1+\delta_{cd}} \int \frac{\mathrm{d}^3p_j}{(2\pi)^3 2E_j} d_q f^{(0)}_j(E_j,T,\mu_q)  \nonumber \\
	&  \times& \int \frac{\mathrm{d}^3p_c}{(2\pi)^3 2E_c}  \int \frac{\mathrm{d}^3p_d}{(2\pi)^3 2E_d} |\bar{\mathcal{M}}|^2 (p_i,p_j,p_c,p_d)\ \nonumber \\  
&  \times &(2\pi)^4 \delta^{(4)}\left(p_i + p_j -p_c -p_d \right) (1-f^{(0)}_{c}) (1-f^{(0)}_{d}) \nonumber \\
	&= &  \frac{1}{2E_i} \sum_{j=q,\bar{q}} \frac{1}{1+\delta_{cd}} \int \frac{\mathrm{d}^3p_j}{(2\pi)^3 2E_j}  d_q f^{(0)}_j(E_j,T,\mu_q)  \nonumber \\
&  \times& \frac{1}{16 \pi \sqrt{s}}\frac{1}{ p_{cm}} \int \mathrm{d}t	|\bar{\mathcal{M}}|^2 (s,t) (1-f^{(0)}_{c}) (1-f^{(0)}_{d}) \nonumber \\
&= & \sum_{j=q,\bar{q}} \int \frac{\mathrm{d}^3p_j}{(2\pi)^3}\ \ d_q f^{(0)}_j(E_j,T,\mu_q) v_{\text{rel}} \sigma_{ij \rightarrow cd}(s,T,\mu_q) ,\nonumber
\eea 
The indices i and j refer to particles in the entrance channel, c and d to those in the exit channel.  $f^{0}_i$  is  the modified Fermi-Dirac distribution function taking into account the Polyakov loop (Eq. \ref{eq:fpol_qbar}). 
$|\bar{\mathcal{M}}|^2$ denotes the matrix element squared averaged over the color and spin of the incoming partons, and summed over those of the final partons. 
$\sqrt{s}$ can be conveniently calculated from the four-vectors of the incoming partons.  
The cross section without the Pauli blocking factors is  
\bea
 \sigma(\sqrt{s})= \int dt \frac{1}{64 \pi s p_{cm}^2}|\bar{\mathcal{M}}|^2.
 \label{xsec}
 \eea
The relative velocity in the c.m. frame is given by
        \begin{align}
        v_{\text{rel}} =\frac{\sqrt{(p_i \cdot p_j)^2-m_i^2 m_j^2}}{E_i E_j}=\frac{p_{cm}\sqrt{s}}{E_i E_j}.
        \label{eq:v_rel_cm}
             \end{align}
$p_{cm}$ is the momenta of the initial $(i,j)$ as well as of the final quarks 
$(c,d)$ in the c.m. frame given by   
     \begin{align}
         p_{cm}=\frac{\sqrt{(s-(m_{i,c}-m_{j,d})^2)(s-(m_{i,c}+m_{j,d})^2)}}{2\sqrt{s}}.
        \label{eq:p_cm}
     \end{align}
     
     \begin{figure}[!h]
 \centering
\center{\includegraphics[width=0.8\linewidth]{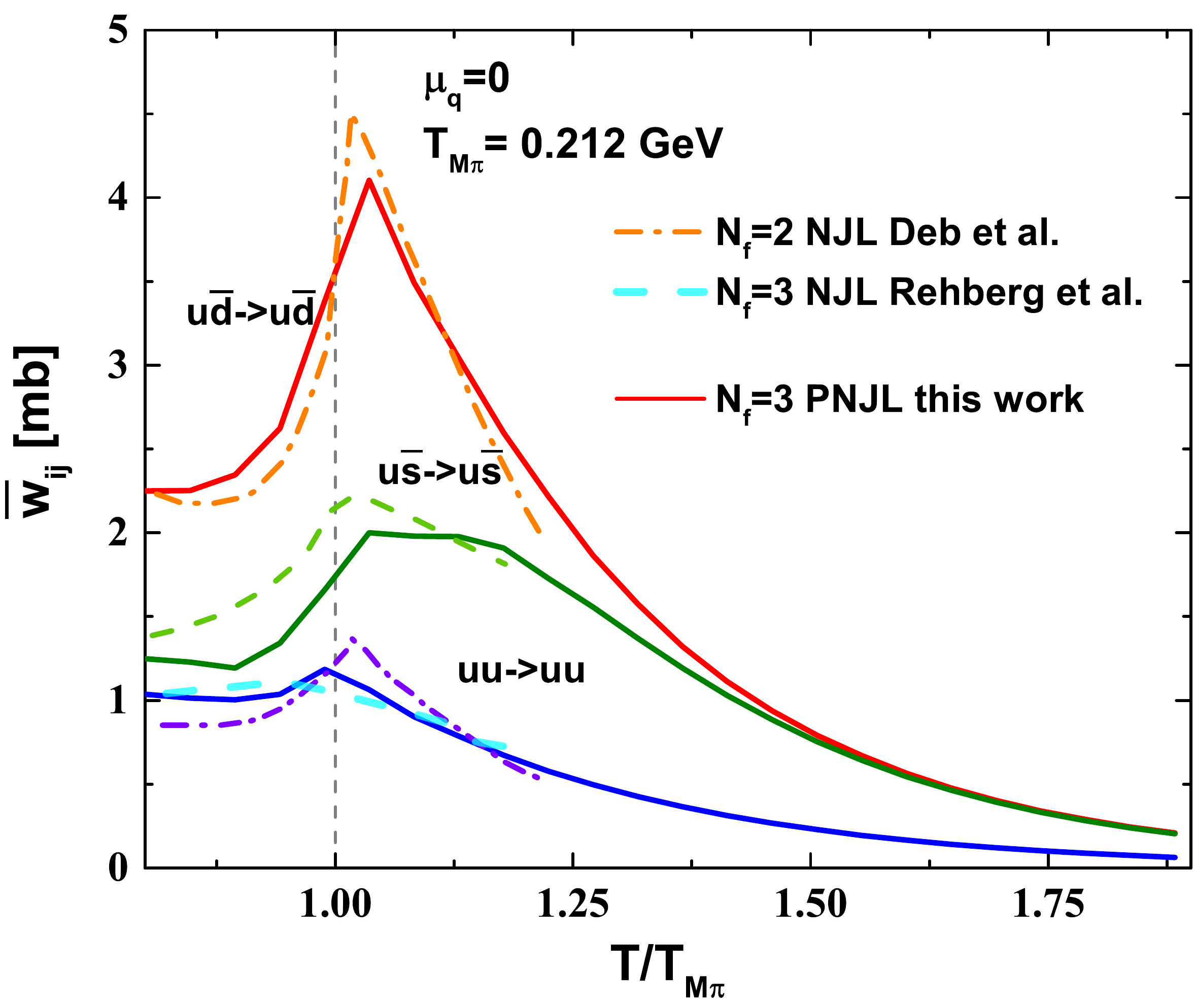} }
  \caption{ Energy averaged transition rates  $\bar w_{ij}(T,\mu_q)$ for different quark-quark(antiquark) scattering processes ($u\overline{d}\rightarrow u\overline{d}$ (red and orange lines), $u\overline{s}\rightarrow u\overline{s}$(green lines), $uu\rightarrow uu$(blue, cyan, and violet lines)) as a function of scaled temperature $T/T_{M\pi}$ for $\mu_q=0$. The solid lines corresponds to the actual results from Eq.~(\ref{rateij}). Green and cyan dashed lines correspond to the results from Ref.~\cite{Rehberg:1996vd}. Orange and violet dash-dotted lines correspond to the estimations from Ref.~\cite{Deb:2016}}
  \label{fig:rate_mu0}
    \end{figure}
    
\begin{figure}[!h]
 \centering
\begin{minipage}[h]{0.8\linewidth}
\center{\includegraphics[width=1\linewidth]{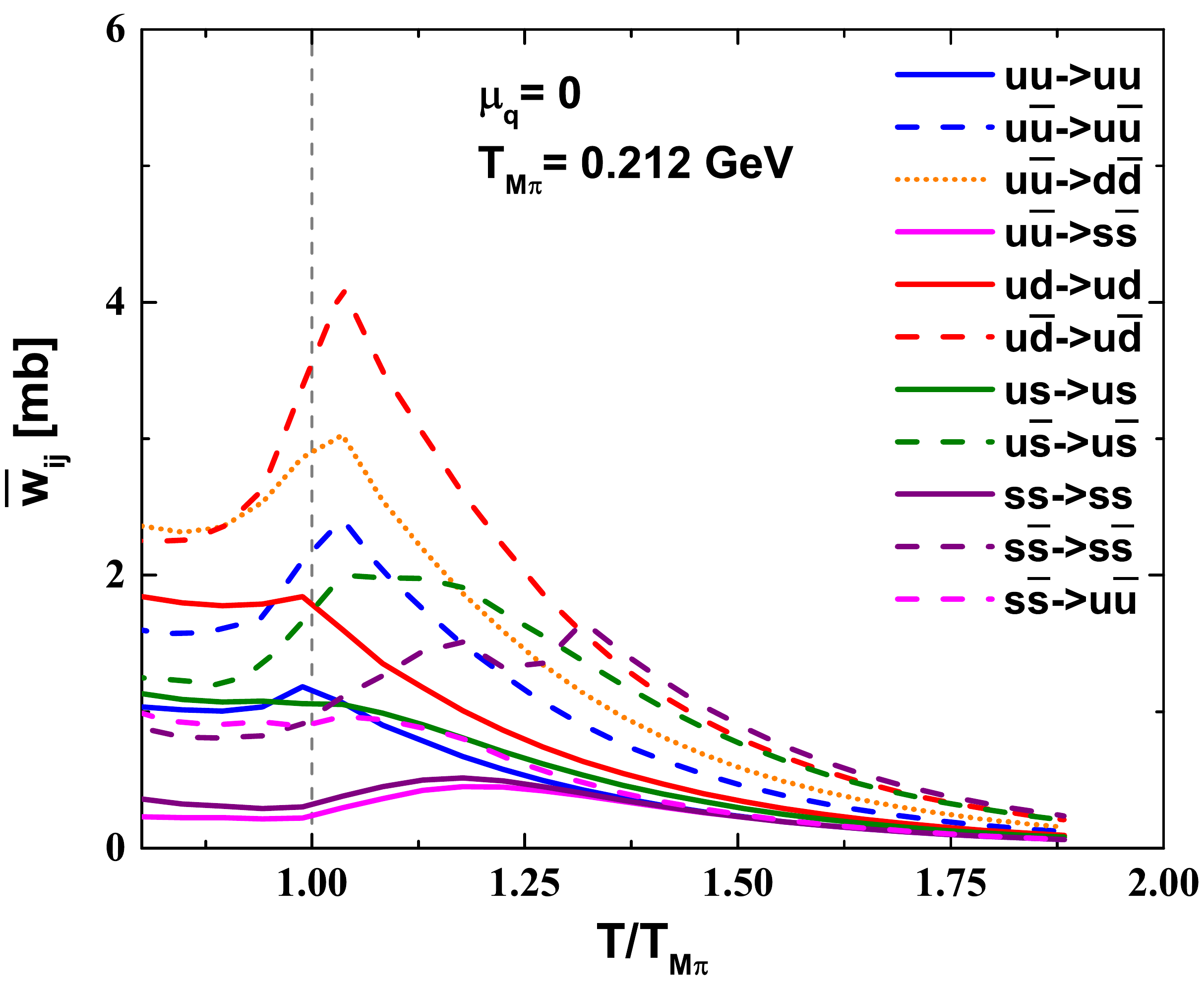} \\ a)}
\end{minipage}
\begin{minipage}[h]{0.8\linewidth}
\center{\includegraphics[width=1\linewidth]{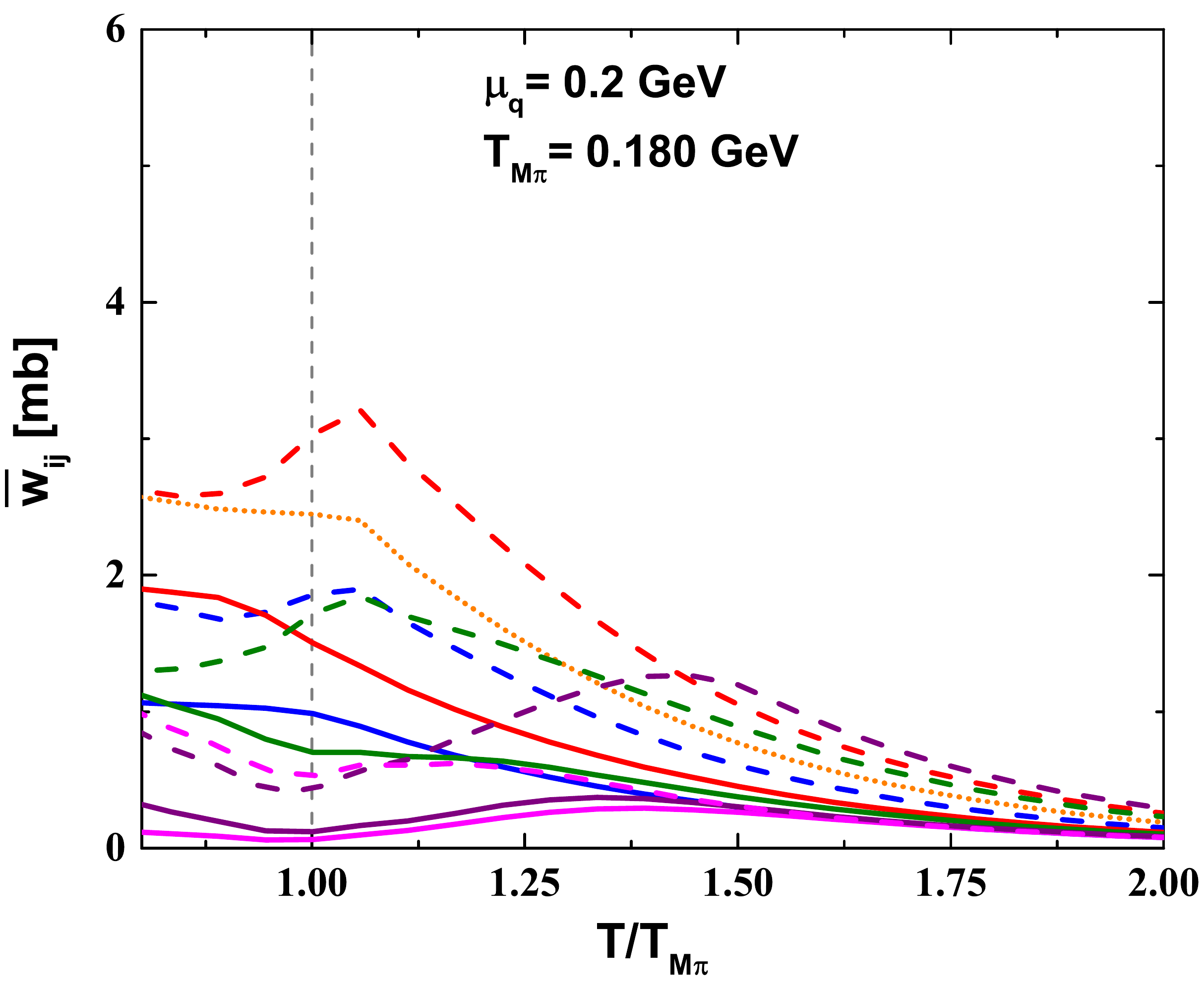} \\ b)}
\end{minipage}
  \caption{Energy averaged transition rates $\bar w_{ij}(T,\mu_q)$ for different quark-quark(antiquark) scattering processes as a function of scaled temperature $T/T_{M\pi}$ 
  for a) $\mu_q=0$ (upper) and b) $\mu_q=0.2$ GeV (lower). The solid and dashed lines correspond to the actual results from Eq.~(\ref{rateij}) for the quark-quark and the quark-antiquark scatterings. }
  \label{fig:rates}
    \end{figure}

The averaged relaxation time can be obtained from the relaxation time of Eq. (\ref{Gamma_on}) by averaging over $p_i$
\be
\tau_{i}^{-1} (T, \mu_q  )  = \frac{1}{n_i(T,\mu_q) } \int \frac{\mathrm{d}^3p_i}{(2\pi)^3}d_qf^{(0)}_{i}  \tau_i^{-1} (p_i,T,\mu_q  ).
\ee

The relaxation time can be expressed via the averaged transition rate $\bar w_{ij}$ defined as:
\bea
 &&\bar w_{ij}=\frac{1}{n_i n_j} \int \frac{\mathrm{d}^3 p_i}{(2\pi)^3} \int \frac{\mathrm{d}^3p_j}{(2\pi)^3}  \label{rateij} \\
&&\times d_q f^{(0)}_i(E_i,T,\mu_q) d_q f^{(0)}_j(E_j,T\mu_q) 
\cdot v_{\text{rel}} \sigma_{ij \rightarrow cd}(s,T,\mu_q). \nonumber  
\eea

We note, that in spite that $\bar w_{ij}$ is called in the literature 'averaged transition rate',  
it has the dimension of a cross section. 
Using  $\bar w_{ij}$ defined by  Eq. (\ref{rateij}), 
the average quark relaxation time is given by 
\cite{Rehberg:1996vd}:
\begin{align}
\tau^{-1}_{i} (T, \mu_q  ) =\sum_{j=q,\bar{q}} n_j(T,\mu_q) \bar w_{ij}  
\label{tau_relax_rateij} 
\end{align}

Fig.~\ref{fig:rate_mu0} illustrates the results of the energy averaged transition rates  $\bar w_{ij}(T,\mu_q)$ for three scattering processes: $u\overline{d}\rightarrow u\overline{d}$ (red and orange lines), $u\overline{s}\rightarrow u\overline{s}$(green lines), $uu\rightarrow uu$(blue, cyan, and violet lines) as a function of scaled temperature $T/T_{M\pi}$ 
( where $T_{M\pi}$ is the Mott temperature) for $\mu_q=0$ from Eq.~(\ref{rateij}) in comparison to the previous NJL results taken from Ref.~\cite{Rehberg:1996vd} (green and cyan dashed lines, $N_f=3$ ) and  Ref.~\cite{Deb:2016} (orange and violet dash-dotted lines, $N_f=2$). Our results are in a good agreement with these NJL results, a small difference arises due to different parameters of the models and different quark masses.
Momentum averaged transition rates $\bar w_{ij}(T,\mu_q)$ for $qq$ (solid lines) and $q \overline{q}$ (dashed lines) scattering channels are presented in Fig.~\ref{fig:rates} as a function of scaled temperature $T/T_{M\pi}$ for a) $\mu_q=0$ and b) $\mu_q=0.2$ GeV.
Near $T_{M\pi}$ the rates $\bar w_{ij}(T,\mu_q)$  have a peak, which is followed by a decrease with increasing temperature. 
While the values of the $q \overline{q}$ rates $\bar w_{q\overline{q}}(T,\mu_q)$ are higher than those of the $qq$ channels, the antiquark densities are smaller than the quark densities at non-zero $\mu_q$ (see Fig.~\ref{fig:dens}).

 
\subsubsection{Method 2 for the quark relaxation time} 

 We continue the estimation of quark relaxation times with an approach which was introduced by Zhuang \cite{Zhuang} for the calculation of the mean free path and then modified by Sasaki \cite{Sasaki} for the evaluation of the relaxation time. 
 It is based on  the 'weighted' thermal averaged  quark-quark and quark-antiquark PNJL cross sections.
	  In the dilute gas approximation the relaxation time for the specie $'i'$ is defined in \cite{Sasaki} as :
\begin{align}
 \tau_i^{-1}(T,\mu_q)=\sum_{j=q,\bar{q}} n_j(T,\mu_q) \overline{ \sigma}_{ij}(T,\mu_q).
 \label{tau_relax_sigij} 
 \end{align}
$\overline{ \sigma}_{ij}(T,\mu_q)$  is the 'weighted' thermal averaged total PNJL scattering cross section
\bea
 \overline{\sigma}_{ij}(T,\mu_q)&=&\int_{s_0}^{s_{max}} \mathrm{d}s \ \sigma_{ij \rightarrow cd}(T,\mu_q,s) \ P(T,\mu_q,s), \nonumber \\ 
 P(T,\mu_q,s)&=&
        C^{'}\int d^3p_i d^3p_j d_q f^{(0)}_i (E_{i},T,\mu_q)d_q f^{(0)}_{j} (E_j,T,\mu_q)   \nonumber \\ 
 & \times& \delta( \sqrt{s}-(E_i+E_{j}))  \ \delta^3 (\vec{p_i}+\vec{p_{j}}) \ v_{rel}. 
\label{eq:sig_ij}
    \eea
Here $P(T,\mu_q,s)$ is the probability of finding a quark-antiquark or quark-quark pair with a center of mass energy $\sqrt{s}$  and a zero  total momentum.
$P(T,\mu_q,s)$ is normalised as 
\bea
\int_{s_0}^{s_{max}} \mathrm{d}s \ P(T,\mu_q,s)=1.
\label{Pnorma}
\eea 
and the relative velocity in the c.m. frame is given by Eq. (\ref{eq:v_rel_cm}).

 For the PNJL results  we use also the modified Fermi-Dirac distribution function defined by Eq.~(\ref{eq:fpol_qbar}).
\begin{figure}[t!]
 \centering 
\begin{minipage}[h]{0.8\linewidth}
\center{\includegraphics[width=1\linewidth]{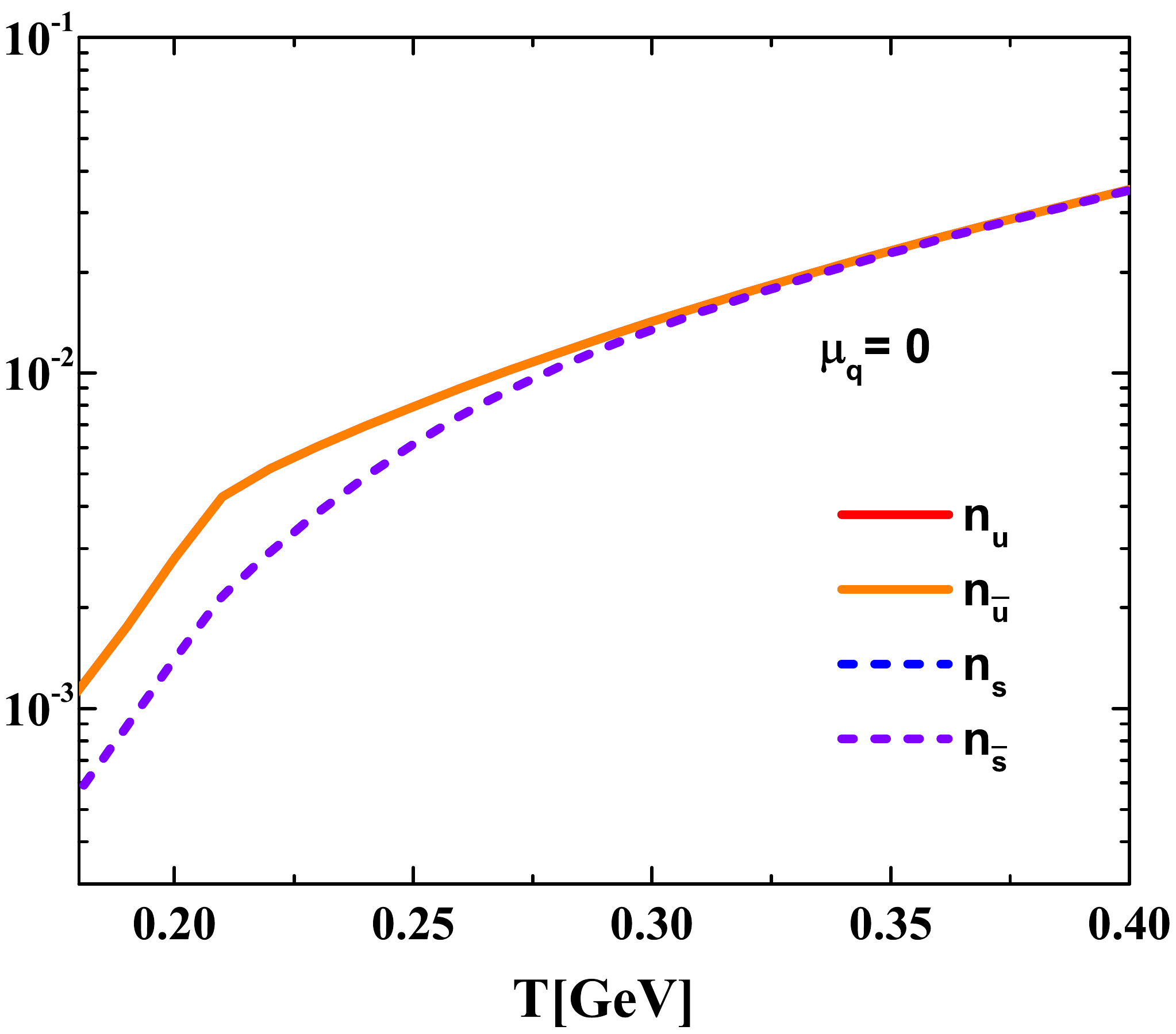} \\ a)}
\end{minipage}
\begin{minipage}[h]{0.8\linewidth}
\center{\includegraphics[width=1\linewidth]{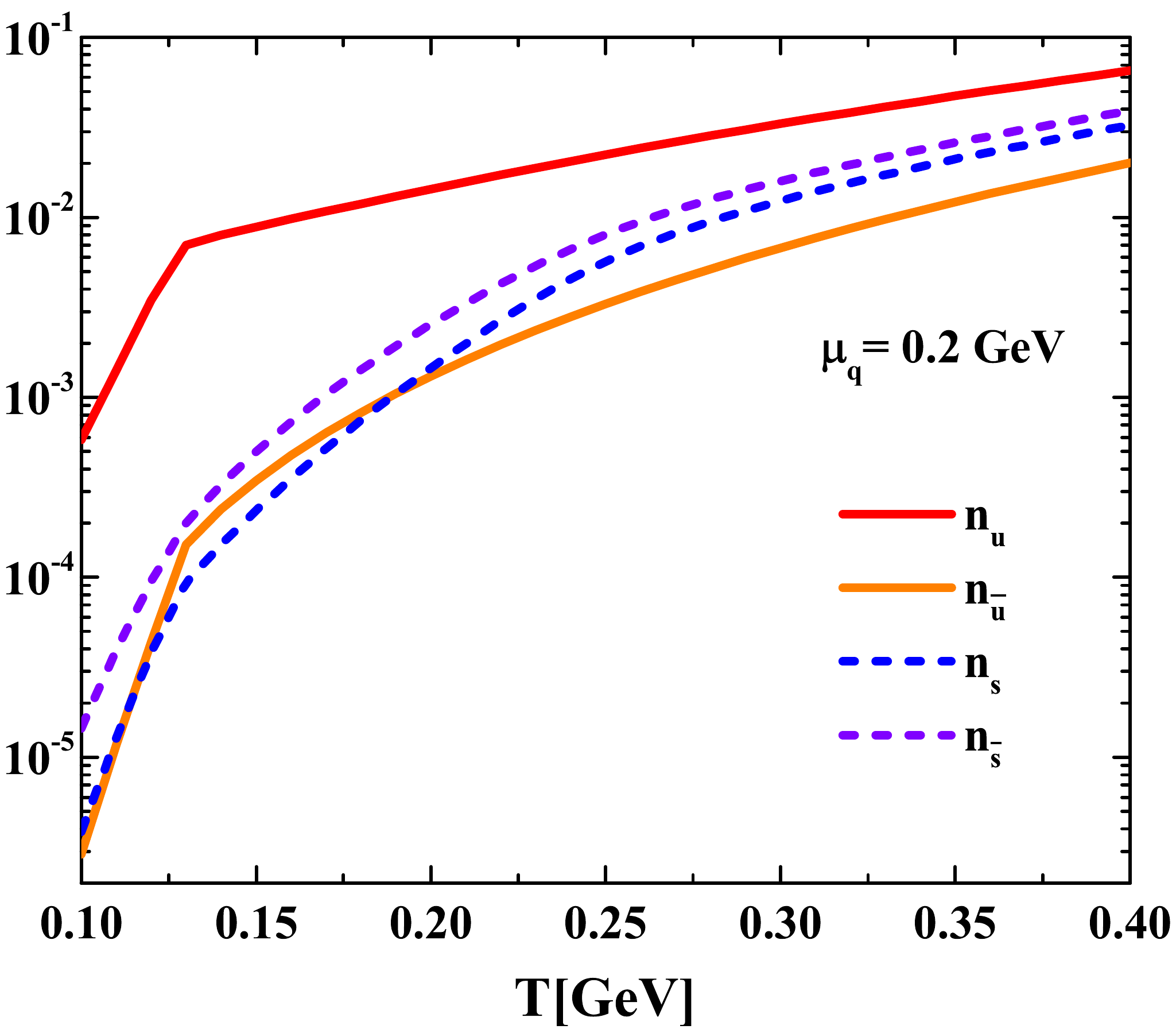} \\ b)}
\end{minipage}
  \caption{Light and strange quarks(antiquarks) densities  $n_i(T,\mu_q) $ with  $f^{\phi}_i$ - modified Fermi distributions from Eq.~(\ref{eq:fpol_qbar}) as a function of temperature for a) $\mu_q=0$ (upper) and b) $\mu_q=0.2$ GeV (lower). 
  The solid orange and red lines correspond to the light quark and antiquark densities, while the dashed blue and violet lines correspond to the strange quark and antiquark densities.}
  \label{fig:dens}
    \end{figure}
Quark densities with the modified Fermi-Dirac distribution functions are shown in Fig.~\ref{fig:dens} as a function of the temperature for a) $\mu_q=0$ and b) $\mu_q=0.2$ GeV.

 The relaxation time for the light quarks is defined as
\bea
\tau_u^{-1}(T,\mu_q)&& = n_u(\overline{ \sigma}_{uu-uu}+\overline{\sigma}_{ud-ud})  
 \label{tauu} \\ 
&& + n_{\bar{u}}(\overline{\sigma}_{u\bar{u}-u\bar{u}}+
   \overline{\sigma}_{u\bar{u}-d\bar{d}} \nonumber \\
&&   +\overline{\sigma}_{u\bar{u}-s\bar{s}}+\overline{\sigma}_{u\bar{d}-u\bar{d}})+ 
   n_s \overline{\sigma}_{us-us}+n_{\bar{s}}\overline{\sigma}_{u\bar{s}-u\bar{s}}. \nonumber 
\eea
 The relaxation time for strange quarks is defined as
     \bea
      \tau_s^{-1} (T,\mu_q) &=&2 n_u \overline{ \sigma}_{us-us} + 2 n_{\bar{u}}\overline{\sigma}_{u\bar{s}-u\bar{s}}  \label{tauu1} \\ 
    &  + & n_s\overline{ \sigma}_{ss-ss}+n_{\bar{s}}(\overline{\sigma}_{s\bar{s}-s\bar{s}}+2 \overline{\sigma}_{s\bar{s}-u\bar{u}}). \nonumber 
     \eea

\begin{figure}[!h]     
\begin{minipage}[h]{0.8\linewidth}
\center{\includegraphics[width=1\linewidth]{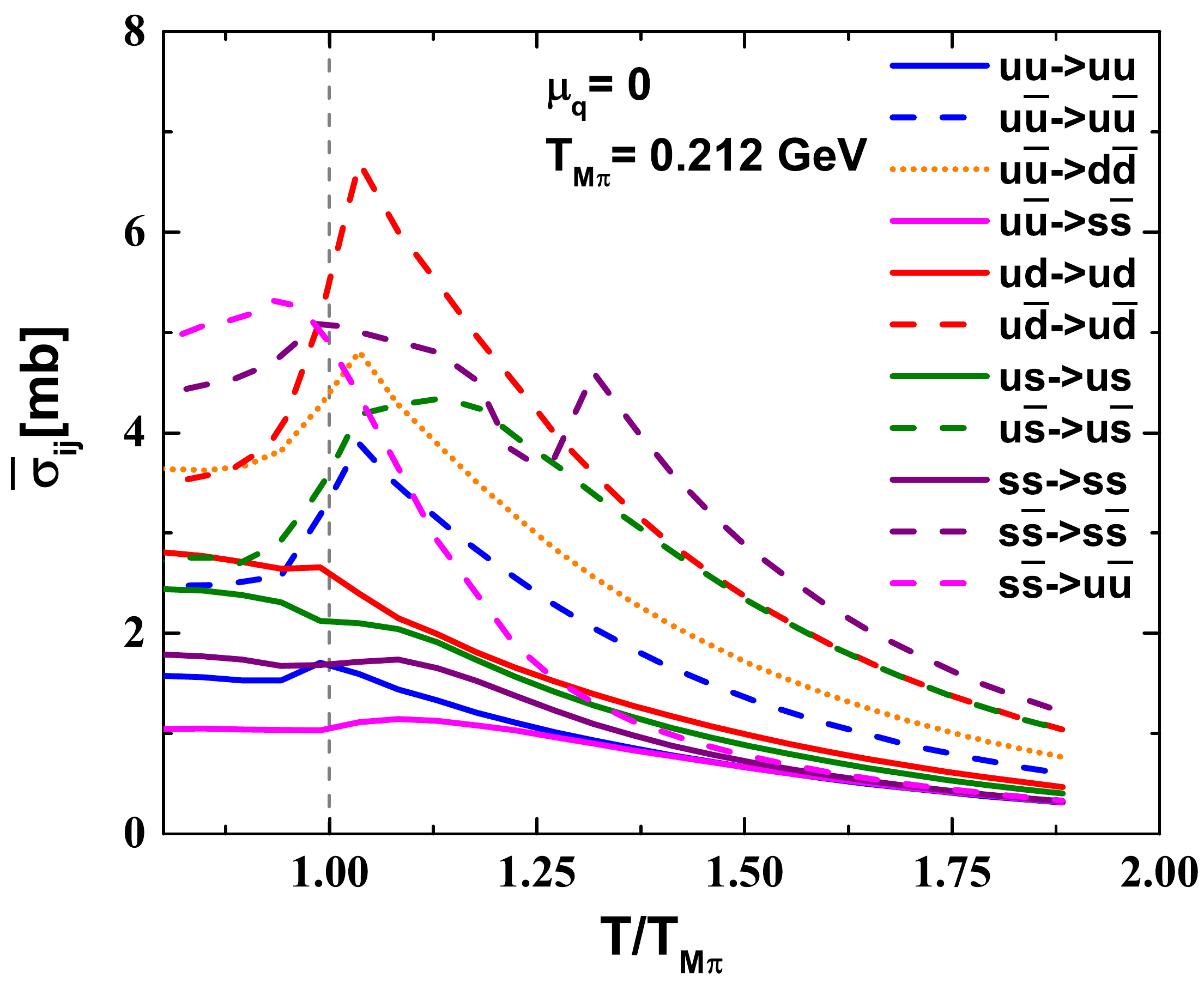} \\ a)}
\end{minipage}
\begin{minipage}[h]{0.8\linewidth}
\center{\includegraphics[width=1\linewidth]{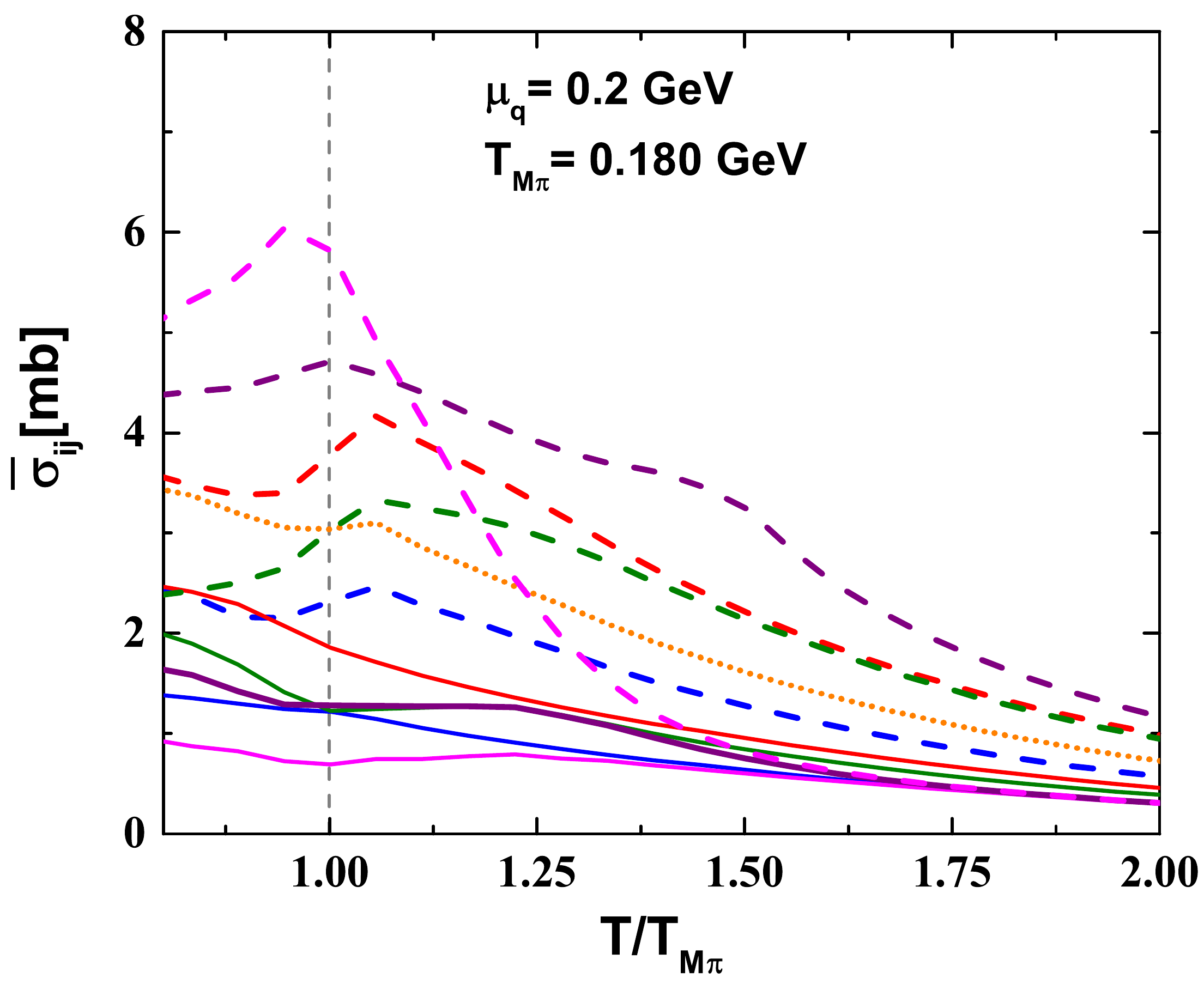} \\ b)}
\end{minipage}
  \caption{'Weighted' thermal averaged total PNJL cross-sections $\overline{\sigma}_{ij}(T,\mu_q)$ from Eq.~(\ref{eq:sig_ij}) as a function of the scaled temperature $T/T_{M \pi}$ for a) $\mu_q=0$ (upper) and b) $\mu_q=0.2$ GeV (lower). }
  \label{fig:sigmaver}
    \end{figure}
 
   Fig.~\ref{fig:sigmaver} shows the 'weighted' thermal averaged PNJL cross sections $\overline{\sigma}_{ij}(T,\mu_q)$ for different scattering processes as a function of the scaled temperature $T/T_{M \pi}$ for a) $\mu_q=0$ and b) $\mu_q=0.2$ GeV.  
   $\overline{\sigma}_{ij}(T,\mu_q)$ shows a peak in the vicinity of the pion Mott temperature $T_{M \pi}$, which is more pronounced for the quark-antiquark $q\bar q$ scattering due to the peak in the cross-sections caused by the s- channel contribution (see discussion in section \ref{sec2} F). 
   Due to this increase of the $q\bar q$ cross sections the 'weighted' thermal averaged cross sections $\overline{\sigma}_{ij}(T,\mu_q)$ for the $q\bar q$ channels dominates over the $qq$ channels. Approaching high temperatures, above the $T_{M \pi}$, the averaged cross sections $\overline{\sigma}_{ij}(T,\mu_q)$ decrease with temperature as it is expected from the behaviour of the total PNJL cross sections presented in the previous section.
    The shape of 'weighted' thermal averaged cross sections for $\mu_q=0$ is similar to the NJL results presented in \cite{Marty:NJL13}, while the absolute values of the PNJL 'weighted' thermal averaged  cross section $\overline{\sigma}_{ij}(T,0)$ are larger due to different model parameters and due to the larger values of the effective quark masses.

Using Eqs. (\ref{rateij}),(\ref{tau_relax_rateij}) and 
(\ref{tau_relax_sigij}),(\ref{eq:sig_ij}) 
one can compare the underlying differences of the two presented methods to calculate  
the quark relaxation time. 
 The first approach is simply an averaging of $v_{rel} \cdot \sigma(\sqrt{s})$ over the momentum of the partons in the entrance channel.
 The second method requires in addition that the sum of the quark momenta in the entrance channel is zero and introduces an additional $\sqrt{s}$ dependence by integrating over 
 $s$ instead of over $\sqrt{s}$.  The first approach does not need any normalization whereas for the second method the normalization covers some of the parameter dependence of  
 $P(s,T,\mu_q)$.

\begin{figure}[h!]
\begin{minipage}[h]{0.8\linewidth}
\center{\includegraphics[width=1\linewidth]{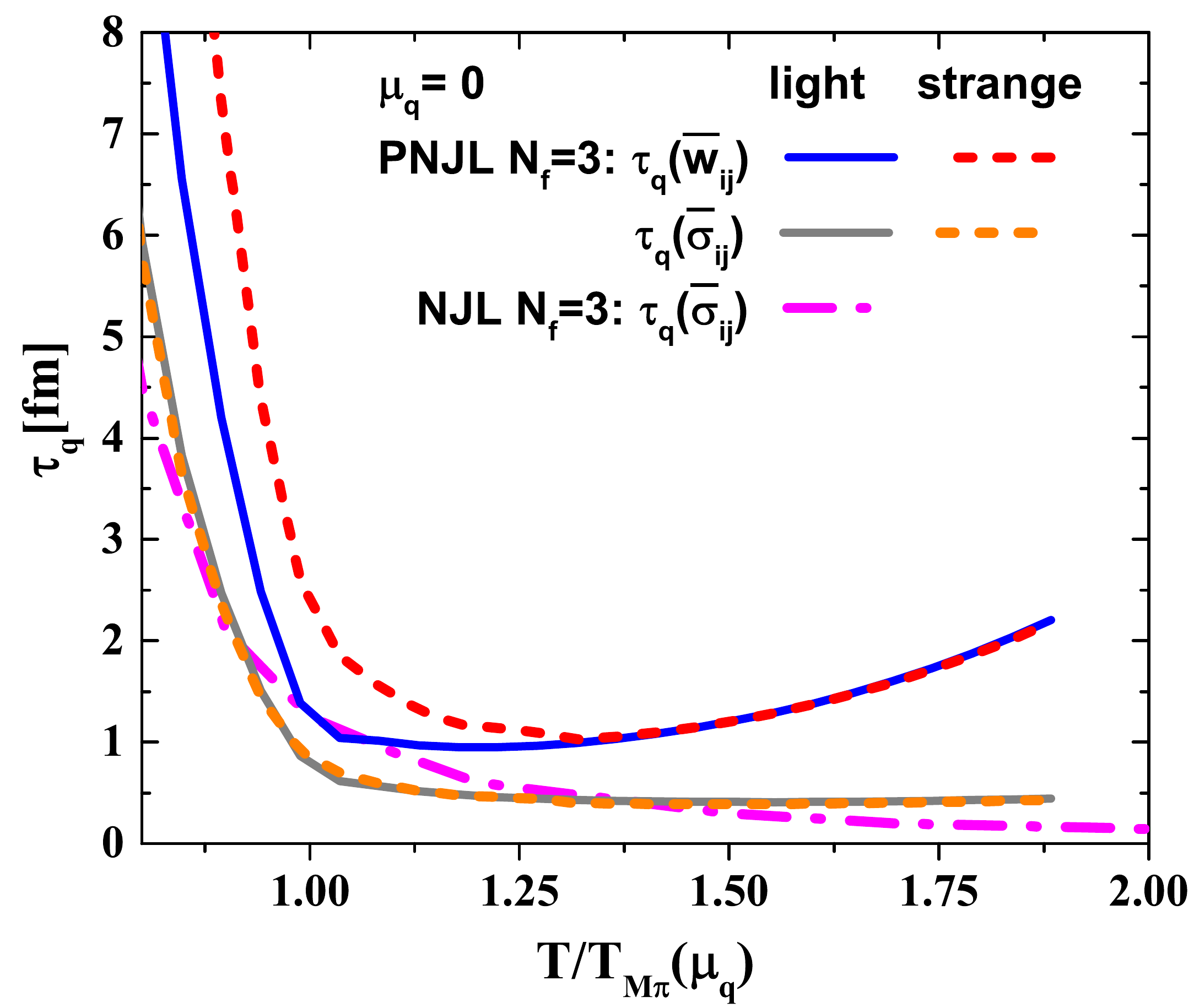} \\ a)}
\end{minipage}
\begin{minipage}[h]{0.85\linewidth}
\center{\includegraphics[width=1\linewidth]{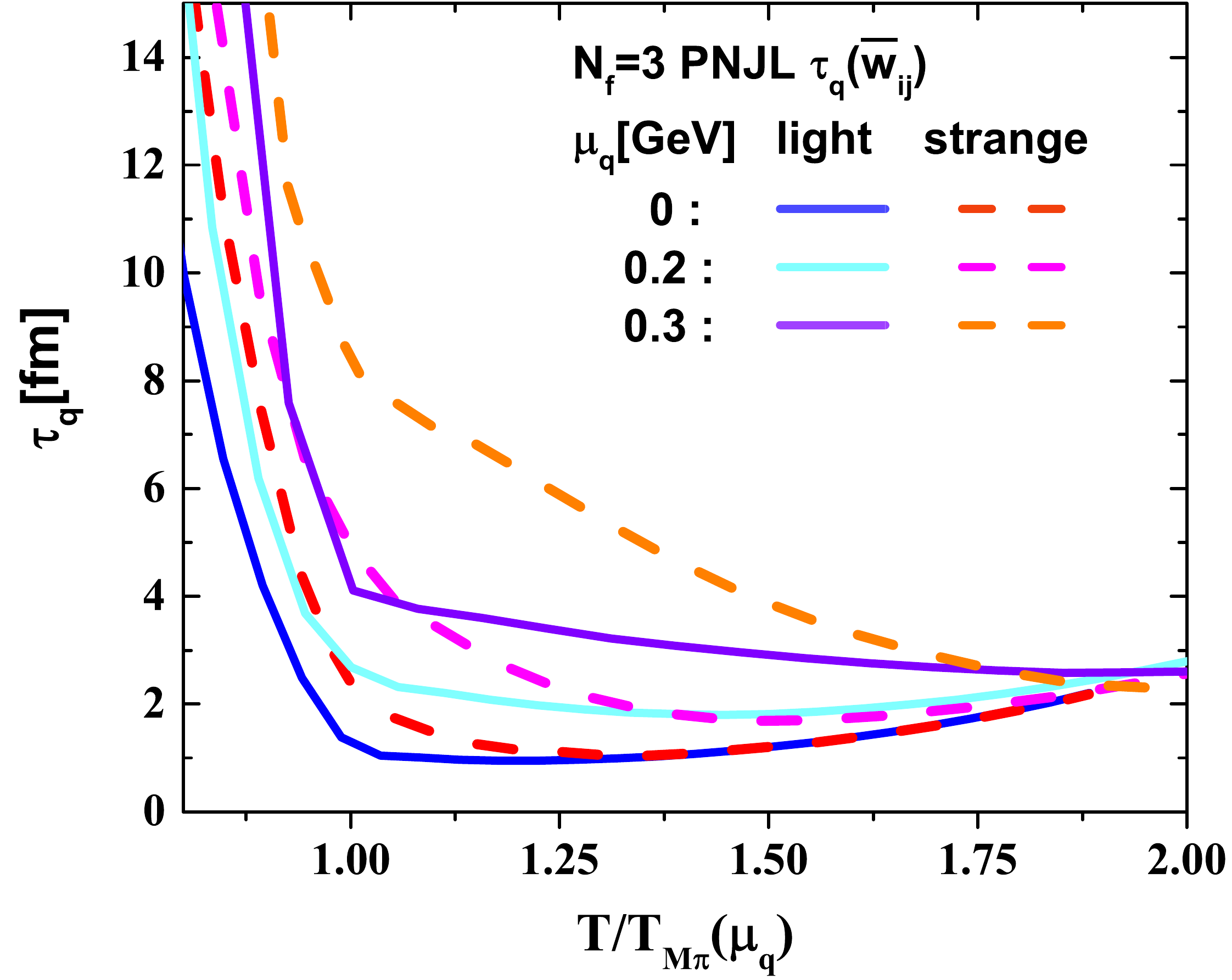} \\ b)}
\end{minipage}
\caption{Relaxation time of  light and strange quarks as a function of the scaled temperature $T/T_{M\pi}(\mu_q)$ for a) $\mu_q=0$ (upper) and b) $\mu_q\geq 0$
(lower). The solid and the dashed lines show the results for the PNJL model using the averaged transition rates $\bar w_{ij}$ (\ref{tau_relax_rateij}) and the 'weighted' thermal averaged cross sections $\bar \sigma_{ij}$ (\ref{tau_relax_sigij}). }
\label{fig:trelax}
\end{figure}   

Fig.~\ref{fig:trelax} a) gives an overview of the relaxation times of  light  and strange quarks as a function of the scaled temperature $T/T_{M\pi}$  and for  $\mu_q=0$. The solid gray and the dashed orange lines correspond to the actual results from Eq.~(\ref{tau_relax_sigij}), where the 'weighted' thermal averaged cross sections $\bar \sigma_{ij}$ are used. The solid blue and the dashed red lines correspond to the results from Eq.~(\ref{tau_relax_rateij}),   where the averaged transition rates $\bar w_{ij}$ are used.

 The  difference  between  the  two  methods  is  most prominently seen at high temperatures. Calculations of the quark relaxation time using the averaged transition rates are more straight-forward since they rely on the relation between the momentum depended relaxation time and the interaction rate. 
 
In addition, we compare the quark relaxation time $\tau_i(\bar \sigma_{ij})$ for the PNJL model with the results for $N_f=3$ NJL model \cite{Marty:NJL13}(dashed magenta line). Our results are in a good agreement with the NJL results except for the  vicinity of $T_{M\pi}$. The light quark relaxation time $\tau_i(\bar \sigma_{ij})$  in this case is about $0.7-0.5$ fm/c in the region $ T_{M\pi} \leq T\leq1.8 T_{M\pi}$. 
 
The $\mu_q$ dependence of the quark relaxation time $\tau_i(\bar w_{ij})$ is shown in Fig.~\ref{fig:trelax} b) for three values of $\mu_q: 0, 0.2$ and $0.3$ GeV. The solid lines correspond to the results for light quarks while the dashed lines correspond to the results for strange quarks. The quark relaxation time is increasing with the chemical potential $\mu_q$ in the region of $T\le 2 T_{M\pi}$. One can see that in the vicinity of the $T_{M\pi}$ the relaxation time for the strange quark is larger than for the light quark. This difference becomes more significant for finite $\mu_q$ due to the difference between the effective mass of light and strange quarks. 

    \subsection{Shear viscosity}
    
 	The most desired transport coefficients are the shear and bulk viscosity. They have been successfully used in the viscous relativistic hydrodynamic description of the QGP bulk dynamics.
 In large systems the shear viscosity and the entropy density scale as $T^3$. Therefore often the specific shear and bulk viscosity are used, the dimensionless ratio of the viscosity to the entropy density. The specific shear viscosity allows to compare the viscosity of liquids at various temperature scales. The main contribution for the viscous description of the QGP comes from the shear viscosity. For this purpose we show the transport coefficients as a function of the scaled temperature $T/T_C$. For the PNJL calculations we use $T_C=T_{M\pi}$
whereas for the other approaches $T_C $ is the temperature of the inflection point. Here we focus on the estimation of the transport coefficients based on the RTA.

The shear viscosity for quarks with medium dependent masses $M(T,\mu_q)$ can be derived using  the Boltzmann  equation in the RTA \cite{Kapusta} through the relaxation time :
\begin{align}
\eta(T,\mu_q)  = \frac{1}{15T} \sum_{i=q,\bar{q}} \int \frac{d^3p}{(2\pi)^3} \frac{\mathbf{p}^4}{E_i^2}    \tau_i(T,\mu_q) 
\cdot  d_q f^{\phi}_i , 
\label{eq:eta_RTA}
\end{align}
where $q(\bar{q})=u,d,s(\bar{u},\bar{d},\bar{s})$,  $\tau_i$ are the relaxation times and $f^{\phi}_i$ are the modified distribution functions, which contain the Polyakov loop contributions.

\begin{figure}[t!]
 \centering
\center{\includegraphics[width=0.8\linewidth]{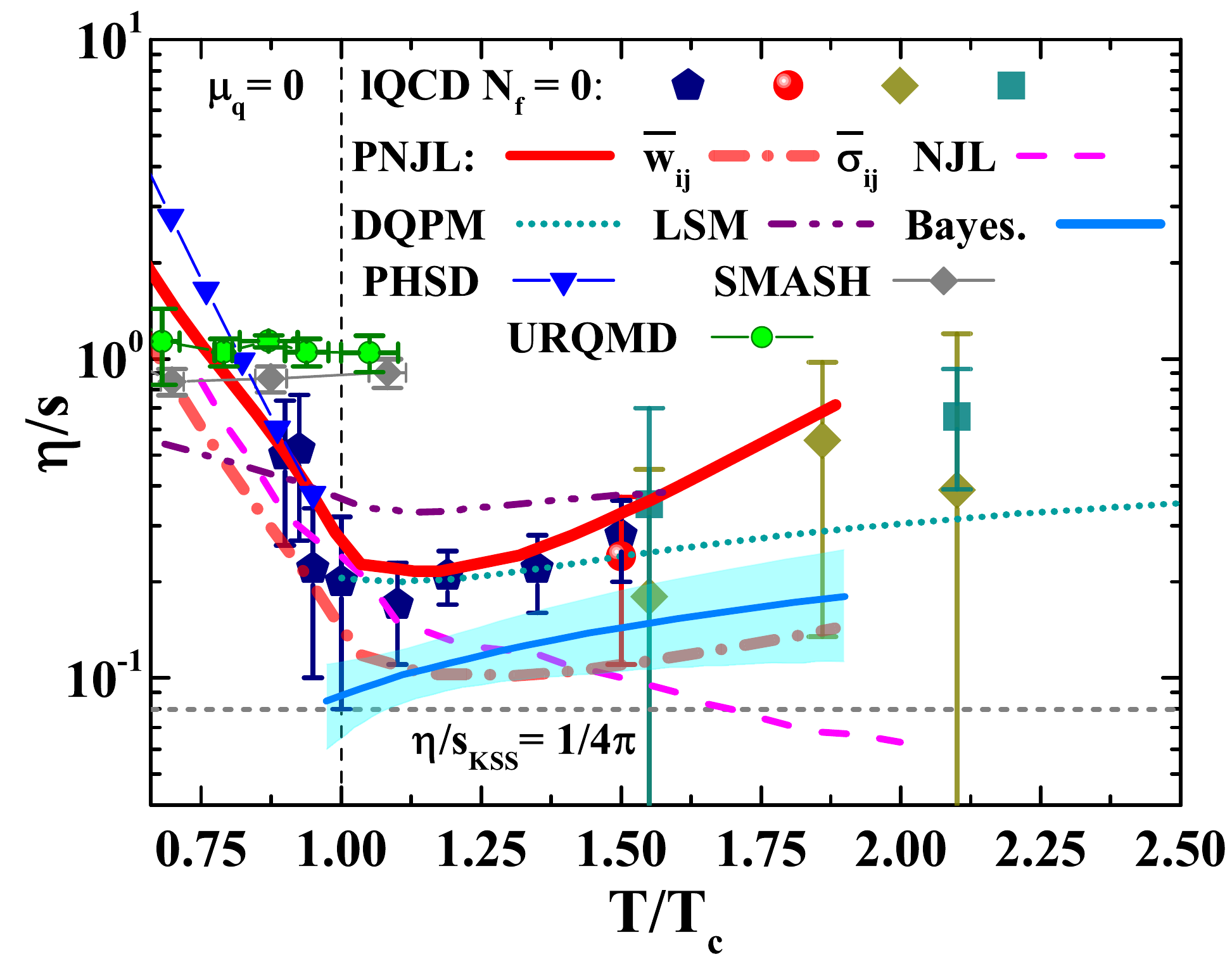} }
  \caption{ Specific shear viscosity $\eta/s$  as a function of scaled temperature $T/T_C$ for $\mu_q=0$. The solid and the dashed red lines show the results of the $\eta/s$ for the PNJL model using the averaged transition rates $\bar w_{ij}$ (\ref{tau_relax_rateij}) and the averaged cross sections $\bar \sigma_{ij}$ (\ref{tau_relax_sigij})  for the evaluation of the relaxation time. We show the estimations from various models: the URQMD \cite{Demir:2008tr} (dotted green line), the PHSD \cite{Ozvenchuk13:kubo} (dotted green line), the SMASH \cite{{Rose:2017bjz}} (dotted green line), the $N_f = 2$ linear sigma model \cite{Heffernan:2020zcf} (dashed doted purple line), the $N_f = 3$
  NJL model \cite{Marty:NJL13} (dashed magenta line),
   DQPM \cite{Soloveva:2019xph} (dotted green line). The dashed gray line demonstrates the Kovtun-Son-Starinets bound \cite{Kovtun:2004} $(\eta/s)_{\rm{KSS}} = 1/(4\pi)$. 
   The symbols show lQCD data for pure SU(3) gauge theory taken from Refs. \cite{Nakamura:2005} (squares and rhombus), \cite{Sakai:2007} (circle), \cite{Astrakhantsev:2017} (pentagons). The solid blue line shows the results from a Bayesian analysis of experimental heavy-ion data \cite{Bernhard:2019bmu}. }
  \label{fig:etas_mu0}
    \end{figure}
    
\begin{figure}[!h]
 \centering
\begin{minipage}[h]{0.8\linewidth}
\center{\includegraphics[width=1\linewidth]{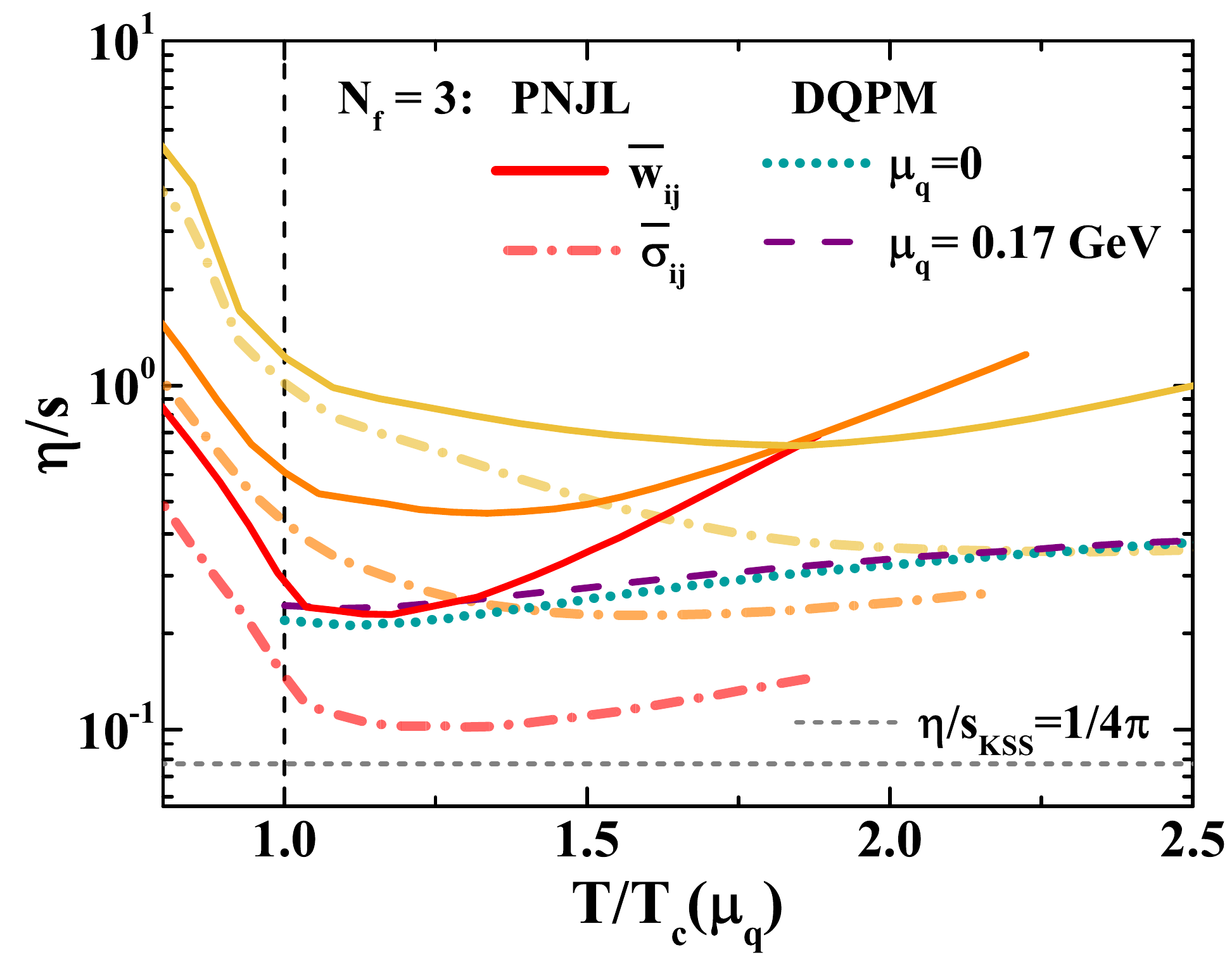} \\ a)}
\end{minipage}
\begin{minipage}[h]{0.8\linewidth}
\center{\includegraphics[width=1.15\linewidth]{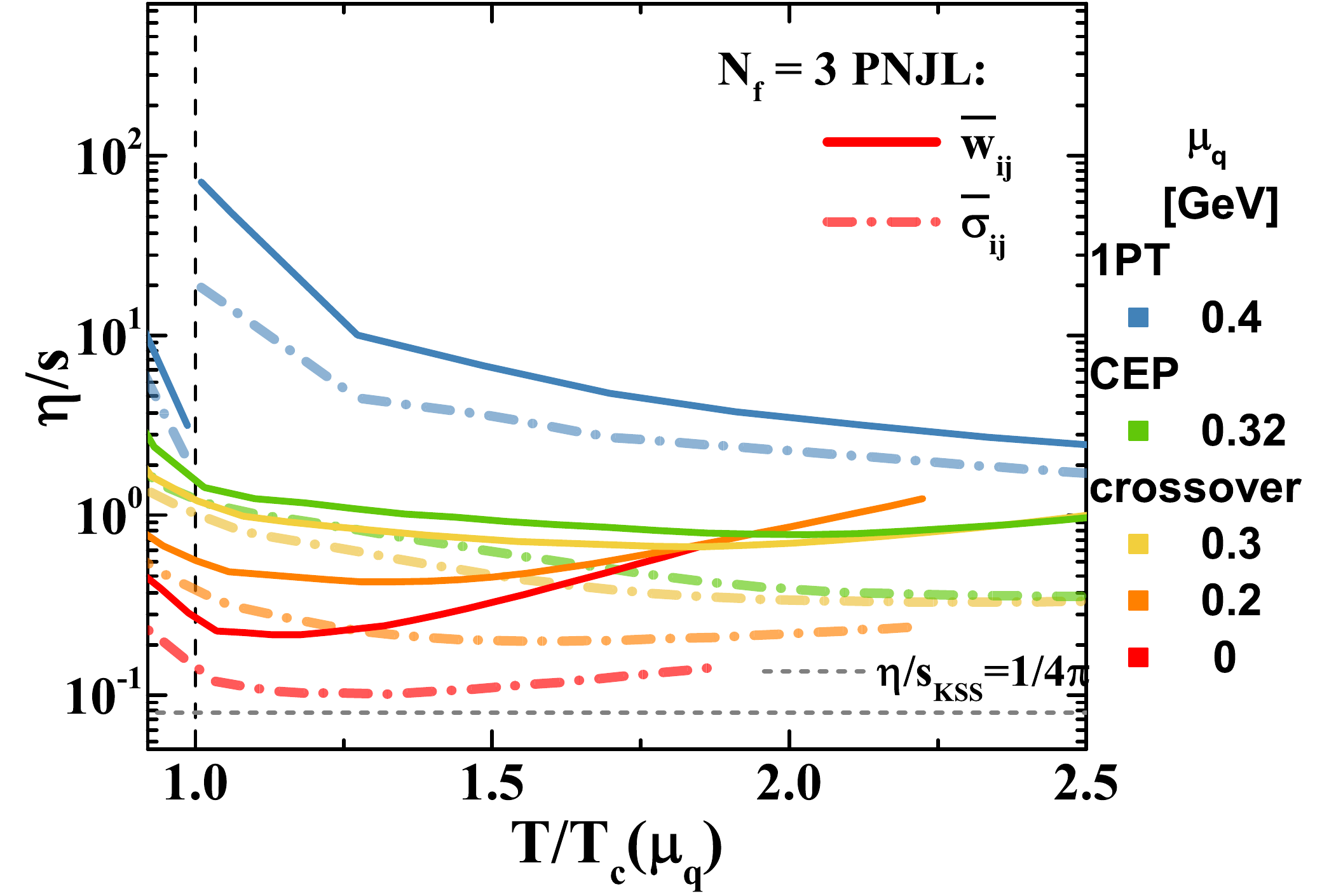} \\ b) }
\end{minipage}
  \caption{ Specific shear viscosity $\eta/s$ as a function of scaled temperature $T/T_C$ for a) (upper) a moderate value of the quark chemical potential $0 \leq \mu_q \leq 0.3$ which corresponds to a crossover phase transition  and b) (lower) whole range of the quark chemical potential $0 \leq \mu_q \leq 0.4$ GeV.  The solid  (dashed) red lines show the PNJL results of $\eta/s$ for the PNJL model using the averaged transition rates $\bar w_{ij}$ (\ref{tau_relax_rateij}) (the averaged cross sections $\bar \sigma_{ij}$ (\ref{tau_relax_sigij}))  for the evaluation of the relaxation time. The dotted green line and dashed purple line correspond to the results from the DQPM \cite{Soloveva:2019xph} for $\mu_q=0$ and $\mu_q=0.17$ GeV. The dashed gray line demonstrates the Kovtun-Son-Starinets bound \cite{Kovtun:2004} $(\eta/s)_{\rm{KSS}} = 1/(4\pi)$. }
  \label{fig:etas_mu}
    \end{figure}
 
Fig.~\ref{fig:etas_mu0} shows the scaled temperature dependence $T/T_C$  of the specific shear viscosity $\eta/s$ for $\mu_q=0$. The solid and dashed red lines show the PNJL results  from Eq. (\ref{eq:eta_RTA}) using the two different estimations of the quark relaxation time: with the averaged transition rates $\bar w_{ij}$ from Eq.~(\ref{tau_relax_rateij}) and with the   'weighted' thermal averaged cross sections $\bar \sigma_{ij}$ from Eq.~(\ref{tau_relax_sigij}). The dashed gray line demonstrates the Kovtun-Son-Starinets bound \cite{Kovtun:2004} $(\eta/s)_{\rm{KSS}} = 1/(4\pi)$, and the symbols show lQCD data for pure SU(3) gauge theory, taken from Refs. \cite{Nakamura:2005} (squares and rhombus), \cite{Sakai:2007} (circle), \cite{Astrakhantsev:2017} (pentagons). The solid blue line presents an estimation of $\eta/s$ from the Bayesian analysis of the experimental heavy-ion data from Ref.~\cite{Bernhard:2019bmu}, which has a similar temperature dependence. The result of $\eta/s$ (using $\bar \sigma_{ij})$ is twice smaller then $\eta/s$ (using $\bar w_{ij})$ due to the different values of the quarks relaxation times. 

We compare the results as well  with those for the $N_f=3$ NJL model from \cite{Marty:NJL13}, where the relaxation time is estimated using Eq.~(\ref{tau_relax_sigij}) and  with the DQPM prediction where the relaxation time is estimated using the on-shell interaction rate described by Eq.~(\ref{Gamma_on}). As expected, $\eta/s$ obtained within the second method is close to the NJL estimation, and differs only at high temperature due to small differences in the cross-sections, while the first method predicts a higher value of $\eta/s$, which is remarkably close to the DQPM results and to the pure SU(3) gauge calculations. 
For the hadronic phase we show the estimations from various transport models:  the URQMD \cite{Demir:2008tr} (dotted green line), the PHSD \cite{Ozvenchuk13:kubo} (dotted green line), the SMASH \cite{Rose:2017bjz} (dotted green line). 
The PNJL results for both methods show a similar temperature dependence in the vicinity of the chiral phase transition. Approaching the phase transition $\eta/s$ has a dip, which is followed by an increase in the high temperature region.  
Later we consider results for non-zero chemical potential, where in the crossover region the DQPM calculations show a very moderate dependence on the chemical potential (for $\mu_u=\mu_s=\mu_B/3$), while the PNJL predictions have a more pronounced $\mu_q$ dependence. As we can see later, for the whole range of the quark chemical potential both methods result in a similar temperature behavior when approaching the chiral phase transition.

Fig.~\ref{fig:etas_mu} a) depicts the specific shear viscosity for moderate values of the quark chemical potential $0 \leq \mu_q \leq 0.3$ GeV where the phase transition is a rapid crossover. We compare our results with the estimations from the DQPM for $\mu_q=0$ (dotted green line) and $\mu_q=0.17$ GeV (dashed violet line). At moderate values of $\mu_q$ the specific shear viscosity shows a  dip after the phase transition, which is vanishing at high values of $\mu_q$ as  can be seen in Fig.~\ref{fig:etas_mu} b). For large $\mu_q$, where the crossover transition turns into the 1st order phase transition (1PT), the specific shear viscosity has a discontinuity near the critical temperature.  In the vicinity of the CEP, for $\mu_q=0.32$ GeV, there is a rather smooth change  of $\eta/s$, which can be seen for the crossover phase transition at $\mu_q=0.3$ GeV. So if one considers only  $\mu_q$ values below the CEP,  
 the temperature dependence of the specific shear viscosity can not point out the position of the CEP. The evolution of the specific shear viscosity with  $\mu_q$ is in qualitative agreement with previous findings made for the $N_f=2$ NJL model in Ref.~\cite{Sasaki}. The numerical  values differ due to the difference in the model parameters, distribution functions and in the NJL entropy density.   

 	\subsection{Electric conductivity}

 As QGP matter consists of  charged constituents it is interesting to estimate the response of the system to an external stationary electric field. This is  described by the electric conductivity $\sigma_0$. Electric conductivity influences the soft photons spectra \cite{Turbide,Akamatsu11,Linnyk13} and it is directly related to their emission rate \cite{Yin}. The electric conductivity can be used for estimation the electromagnetic fields produced in HICs\cite{Oliva:2020mfr}. 

The electric conductivity $\sigma_0$ of quarks with the effective masses $M(T,\mu_q)$ can be evaluated by using the relaxation time approximation (see Ref.~\cite{Thakur}):
\begin{align}
    \sigma_0(T,\mu_q) = \frac{e^2}{3T} \sum_{i=q,\bar{q}}  q_i^2 \int \frac{d^3p}{(2\pi)^3} \frac{\mathbf{p}^2}{E_i^2}     
 \cdot \tau_i(T,\mu_q) \ d_q \ f^{\phi}_i , 
\label{eq:sigm}
\end{align}
where $e^2=4 \pi \alpha_{em} $, $q_i=+2/3(u),-1/3(d),-1/3(s)$ are the quark electric charges, 
$d_q = 2N_c = 6$ is degeneracy factor for spin and color in case of quarks and antiquarks, $\tau_i$ are their relaxation times, while $f^{\phi}_i$ denote the modified distribution functions for quark and antiquarks. In these formulas we deal with quarks and antiquarks of $N_f=3$ flavours. Each quark has a contribution proportional to its charge squared. While viscosities have in general a gluonic contribution, the electric conductivity contains only a quark contribution. 
%
 	\begin{figure}[!th]
 \centering
\center{\includegraphics[width=0.8\linewidth]{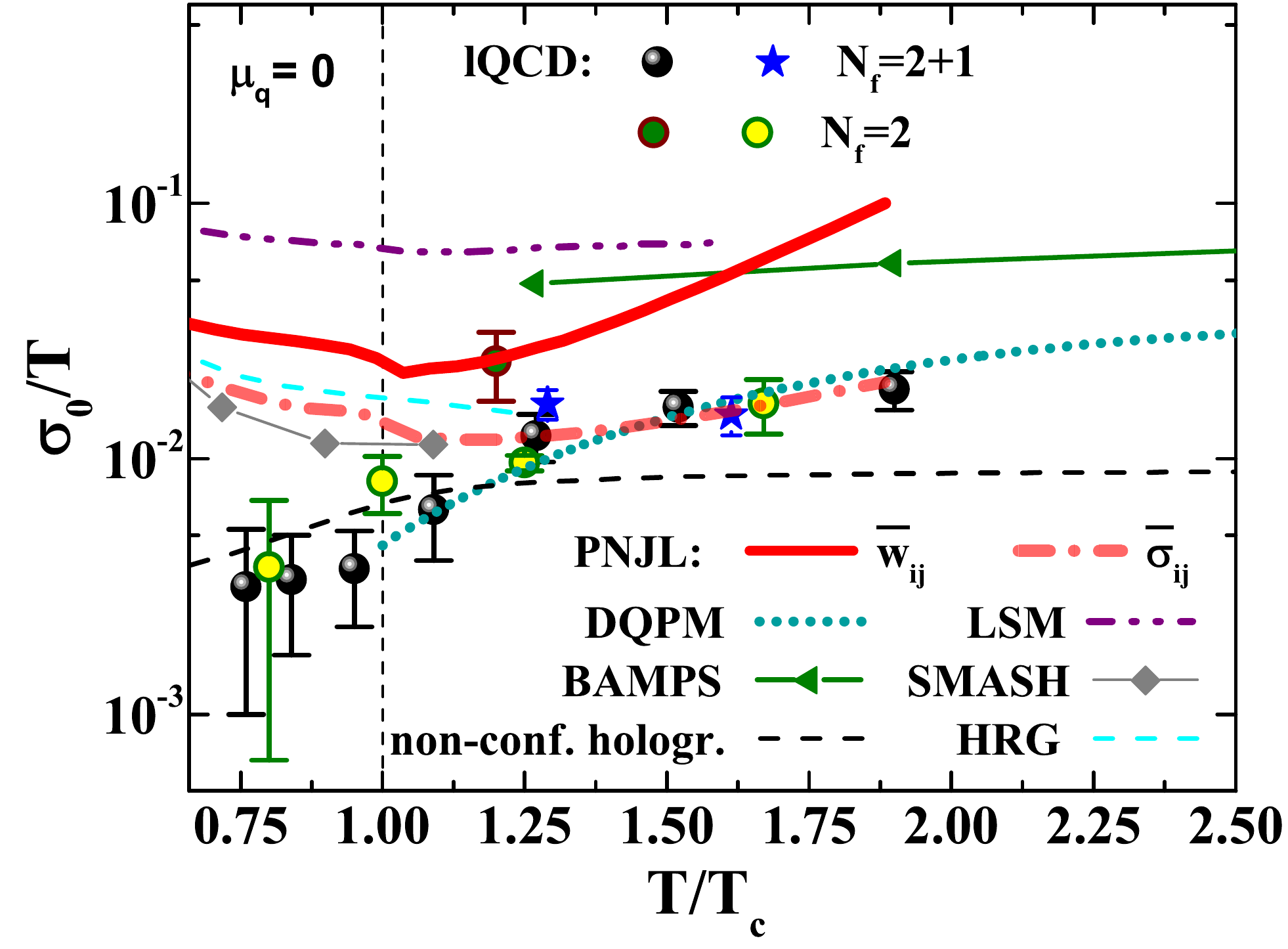} }
  \caption{Ratio of electric conductivity to temperature $\sigma_0/T$ from 
  Eq.~(\protect\ref{eq:sigm}) as a function of the scaled temperature $T/T_C$ for $\mu_q=0$. The solid (dashed) red lines show the PNJL results of  $\sigma_0/T$ for the PNJL model using the averaged transition rates $\bar w_{ij}$ (\ref{tau_relax_rateij}) (the 'weighted' thermal averaged cross sections $\bar \sigma_{ij}$ (\ref{tau_relax_sigij})) for the evaluation of the relaxation time. The symbols display lQCD data for $N_f=2$ taken from Refs.~\cite{Brandt13,Brandt12,Brandt16} (red circles with brown borders), (yellow circles with green borders) and for $N_f=2+1$ taken from Refs.~\cite{Aarts13,Aarts15} (spheres) and from Ref.~\cite{Astrakhantsev:2019zkr} (blue stars). We compare to predictions from the various models: the kinetic partonic cascade model BAMPS  \cite{Greif:2014oia} (the dark-green solid line with triangles), the non-conformal holographic EMD model \cite{Rougemont:hologr} (dashed black line), the DQPM \cite{Soloveva:2019xph} (dotted green line), and below $T_c= 0.158$ GeV we show evaluations from hadronic models: the HRG model within the Chapman-Enskog expansion of the BU \cite{Greif18,Fotakis:2019nbq} (dashed cyan line), the $N_f = 2$ linear sigma model \cite{Heffernan:2020zcf} (dashed doted purple line), the SMASH \cite{Hammelmann:2018ath,Rose:2020sjv} (solid grey line with squares).  }
  \label{fig:cond_mu0}
    \end{figure}

    \begin{figure}[!h]
 \centering
\begin{minipage}[h]{0.8\linewidth}
\center{\includegraphics[width=1\linewidth]{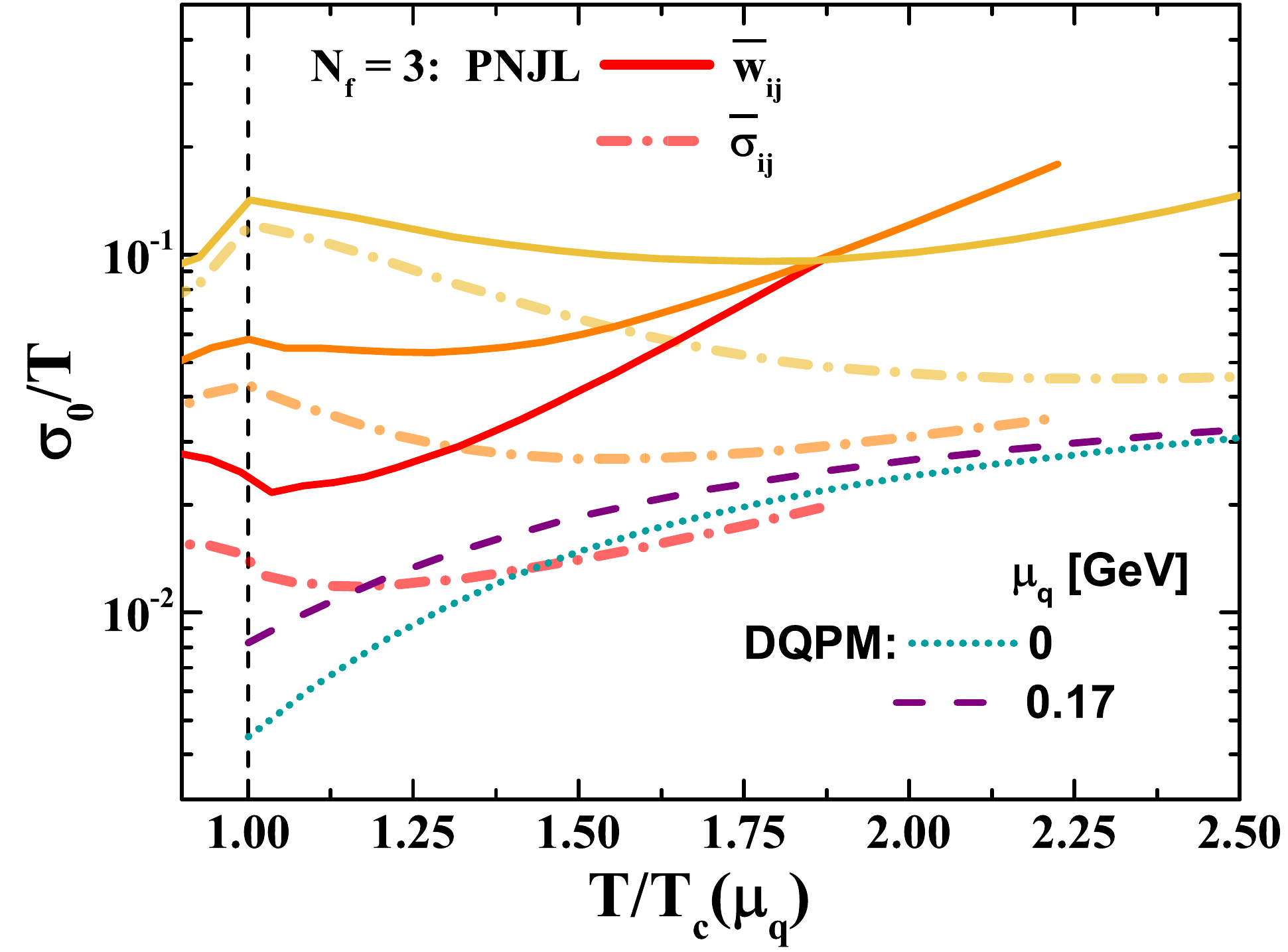} \\ a)}
\end{minipage}
\begin{minipage}[h]{0.8\linewidth}
\center{\includegraphics[width=1.15\linewidth]{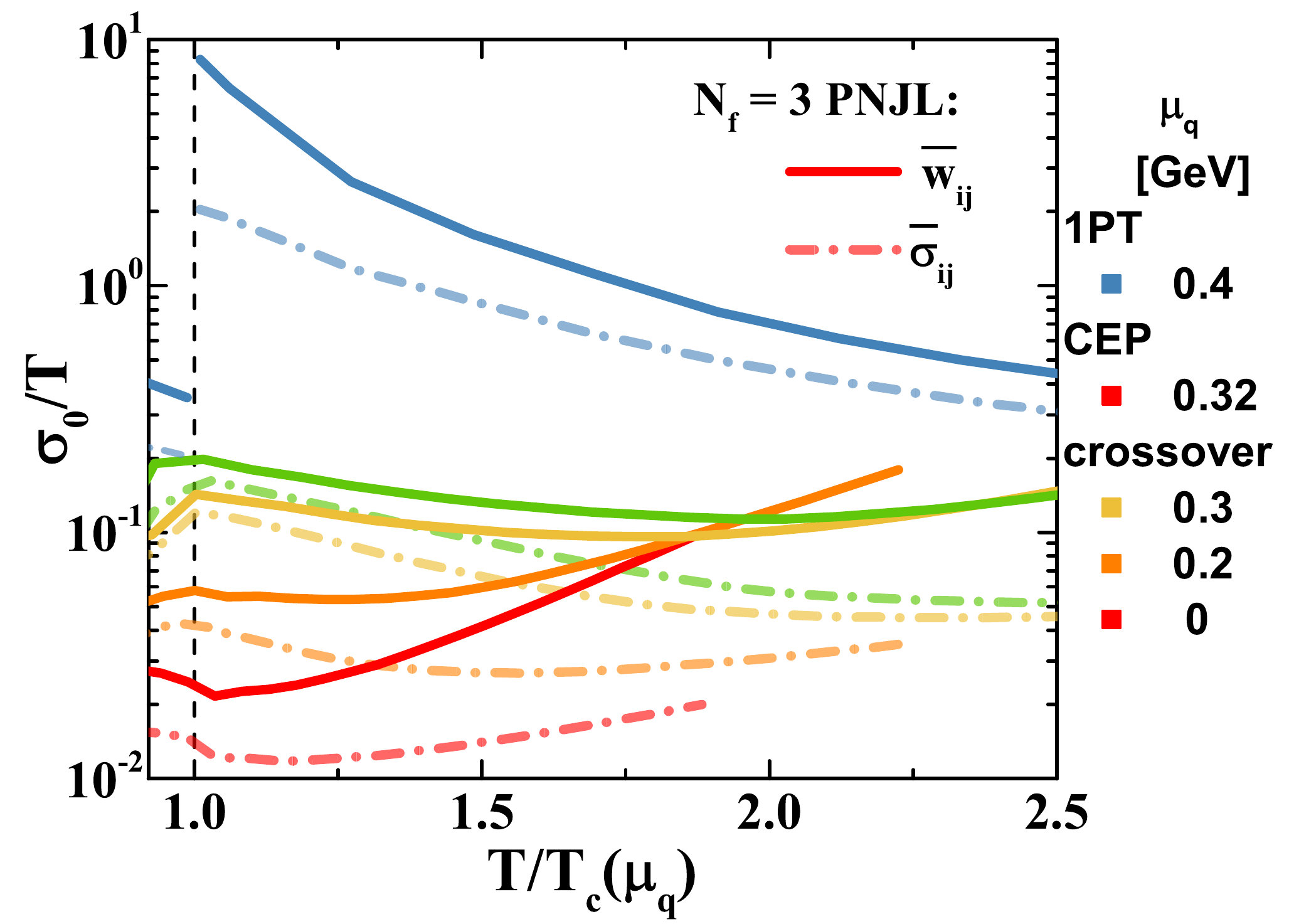} \\ b) }
\end{minipage}
  \caption{ Ratio of the electric conductivity to the temperature $\sigma_0/T$ as a function of the scaled temperature $T/T_C$ for a) $\mu_q=0$ (upper) and 
  b) $\mu_q\geq0$ (lower). The solid (dashed) red lines show the PNJL results of  $\sigma_0/T$  for the PNJL model using the averaged transition rates $\bar w_{ij}$ (\ref{tau_relax_rateij}) (the 'weighted' thermal averaged cross sections $\bar \sigma_{ij}$ (\ref{tau_relax_sigij})) for the evaluation of the relaxation time. The dotted green line and dashed purple line correspond to the results from the DQPM \cite{Soloveva:2019xph} for $\mu_q=0$ and $\mu_q=0.17$ GeV. }
  \label{fig:cond_mu}
    \end{figure}
The PNJL results for the dimensionless ratio of electric conductivity to temperature $\sigma_0/T$ for $\mu_q=0$ are presented in Fig.~\ref{fig:cond_mu0} 
for both methods of the calculation of the quark relaxation time as solid and dashed red lines. We compare the PNJL results to the various estimations from the literature: lQCD data for $N_f=2$ taken from Refs.~\cite{Brandt13,Brandt12,Brandt16} (red circles with brown borders), (yellow circles with green borders) and for $N_f=2+1$ taken from Refs.~\cite{Aarts13,Aarts15} (spheres) and from Ref.~\cite{Astrakhantsev:2019zkr} (blue stars);  the kinetic partonic cascade model BAMPS  \cite{Greif:2014oia} (the dark-green solid line with triangles), the non-conformal holographic EMD model \cite{Rougemont:hologr} (dashed black line), the DQPM \cite{Soloveva:2019xph} (dotted green line), and below $T_c = 0.158$ GeV we show evaluations from hadronic models:  the HRG model within the Chapman-Enskog expansion of the BU \cite{Greif18,Fotakis:2019nbq} (dashed cyan line), the $N_f = 2$ linear sigma model \cite{Heffernan:2020zcf} (dashed doted purple line), the SMASH \cite{Hammelmann:2018ath,Rose:2020sjv} (solid grey line with squares).
The PNJL results for the both methods of the estimation of quark relaxation time have a similar increase with temperature, which is mainly a consequence of  the increase of the quark densities with temperature. The temperature dependence is in agreement with the predictions from the DQPM \cite{Soloveva:2019xph}, despite the differences in the effective masses.

The chemical potential dependence is shown in Fig.~\ref{fig:cond_mu} a) for moderate values of $\mu_q$ and b) for the whole range of $\mu_q$.
\begin{figure}[!h]
 \centering
\center{\includegraphics[width=0.9\linewidth]{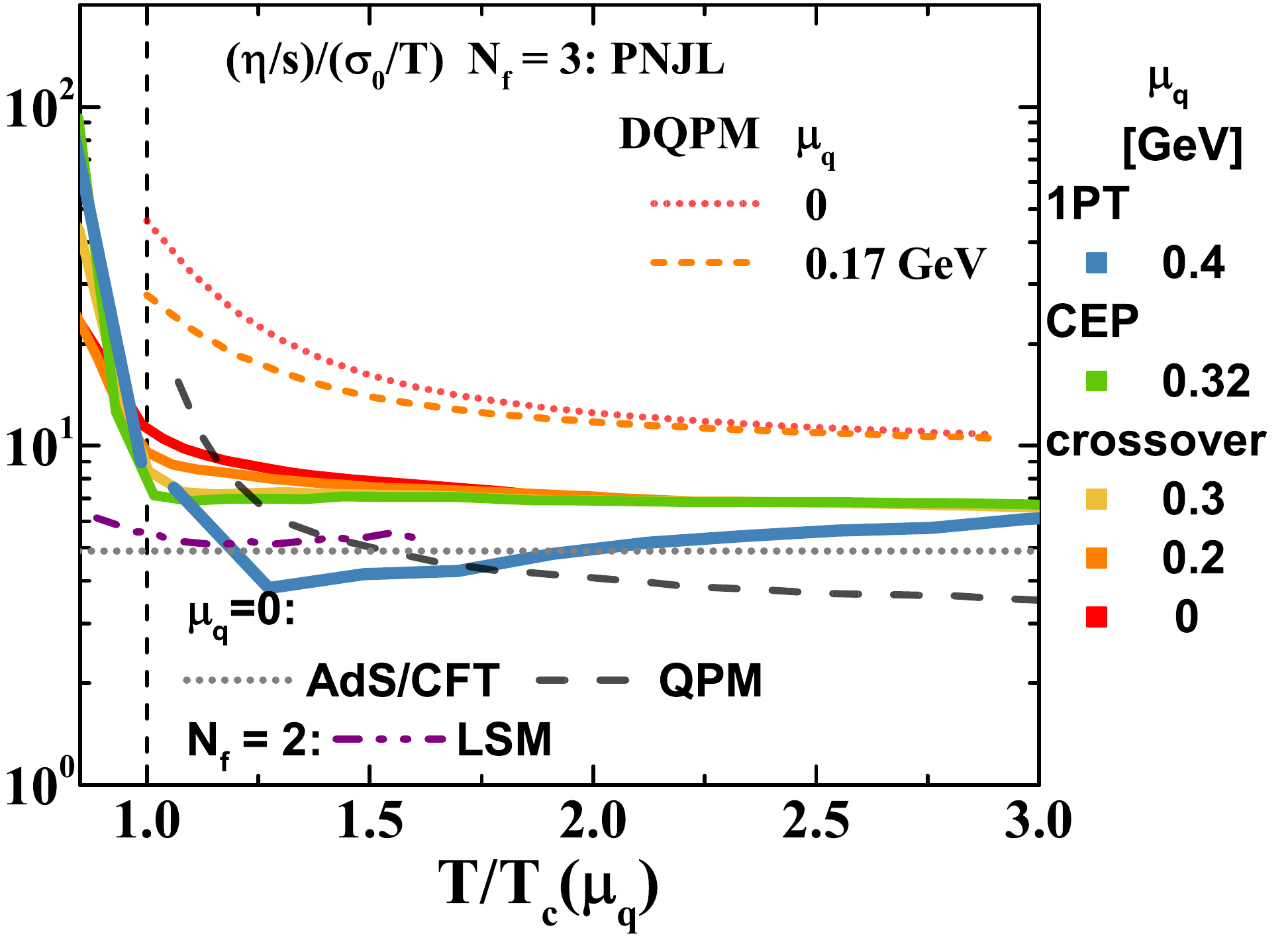} }
  \caption{ Ratio of specific shear viscosity $\eta/s$ to the scaled electric conductivity $\sigma_0/T$  as a function of scaled temperature $T/T_C$ for $\mu_q\geq 0$. For $\mu_q=0$ we show the estimations from various models: the QP model~\cite{Puglisi:2014pda} (dashed grey line), the DQPM \cite{Soloveva:2019xph} (dotted red line), the AdS/CFT ~\cite{Kovtun:2004,CaronHuot:2006te} (dotted grey line), the $N_f =2$ linear sigma model \cite{Heffernan:2020zcf} (dashed doted purple line).}
  \label{fig:etavscond_mu}
    \end{figure}
As the specific shear viscosity, also the electric conductivity has a discontinuity at  the 1st order phase transition (and hence for $\mu_q > 0.32$ ). At lower chemical potentials, where the transition is a crossover,  $\sigma_0/T$ is a continuous function of the temperature. 
Starting from  $\mu_q=0$,  with increasing $\mu_q$, $\sigma_0/T$  has first a dip  approaching the phase transition temperature, which, for a moderate value of $\mu_q=0.2-0.3$ GeV, turns into a hump before at  $\mu_q=0.4$ GeV, where the phase transition is of 1st order, it  shows a discontinuity.
For low values of $\mu_q$  and above  the chiral phase transition, $T \leq 2 T_C$,   $\sigma_0/T$   is raising with $\mu_q$, which is in agreement with the DQPM estimations \cite{Soloveva:2019xph}, and predictions from the holographic calculations in Ref.~\cite{Rougemont:hologr}.

It is interesting to compare  the momentum diffusion, described by the specific shear viscosity $\eta/s$,  and the charge diffusion, described by the scaled electric conductivity $\sigma_0/T=\kappa_Q/T^2$($\kappa_Q$ is the charge diffusion coefficient) by calculating  the ratio $\frac{\eta/s}{\sigma_0/T}$. This ratio is less dependent on the approximations made for the evaluation of the quark cross-sections or quark relaxation times. The ratio $\frac{\eta/s}{\sigma_0/T}$  is presented in Fig.~\ref{fig:etavscond_mu} as a function of scaled temperature $T/T_C$ for a range of quark chemical potential $\mu_q\geq 0$. 

For $\mu_q=0$ we compare our results with the predictions from the AdS/CFT (grey dotted line)~\cite{Kovtun:2004,CaronHuot:2006te}, evaluations from the $N_f =2 $ linear sigma model \cite{Heffernan:2020zcf} (dashed doted purple line), the DQPM predictions \cite{Soloveva:2019xph}(green dash-dotted line) and estimation made in the quasi-particle (QP) model (blue dashed line)~\cite{Puglisi:2014pda},  where quarks and gluons are on-shell particles, and the coupling constant has a one-loop pQCD ansatz, which results in the higher parton masses compare to the DQPM masses. 
Note that the QPM has a higher value of the electric conductivity compared to the DQPM and lQCD results. 
For the PNJL calculations we see for $\mu_q$ values, where the theory shows a crossover, below $T_C$  a strong decrease of this ratio with temperature, whereas above $T_C$ the ratio flattens out.
For $\mu_q$ values where a first order phase transition is observed we see also for this ratio the discontinuity which we already observed for the viscosity and the electric conductivity. \\
 It has to be mentioned that the 3-dimensional mean field models  are conceptually not accurate near the critical point and the first order phase transition \cite{Goldenfeld:1992qy}. They are  built on the anzatz that the fluctuations are small compared to the average value, while approaching the critical point the correlation length becomes large and diverges at the critical point. The feature of the PNJL and NJL models regarding the static critical exponents, the size of the critical region and the influence of the Polyakov loop have already been studied in Refs \cite{Costa:2007ie,Costa:2011fh,Schaefer:2011ex}.

 In the vicinity of the critical region one has to consider additional critical contributions governed by the dynamics of the fluctuations associated with the CEP. The dynamical universality class of the  QCD  critical  point  is  argued  to  be  that  of  the H-model \cite{Fujii:2003,Son:2004iv,Fujii:2004jt} according to the classification of dynamical critical phenomena by Hohenberg and Halperin  \cite{Hohenberg:1977ym}. Whereas in the vicinity of the CEP the shear viscosity has a mild divergence in the critical region, the bulk viscosity has a more pronounced divergence \cite{Kadanoff:1968zz,Hohenberg:1977ym,Son:2004iv,Onuki:2002}: $\eta \sim \xi_T ^{Z_\eta} (Z_\eta \approx 1/19)$,  $\zeta \sim \xi_T^{Z_\zeta}(Z_\zeta \approx 3)$, and electric conductivity  diverges as  $\sigma_Q \sim \frac{1}{\xi_T}$, where $\xi_T \sim (T-T_C)^\nu$ is the thermal correlation length, with $\nu$ being the static critical exponent. The specific bulk and shear viscosities have been considered near the CEP and the 1st order phase transition for the $N_f = 2$ NJL model in the previous study \cite{Sasaki}. Therefore the presented results can qualitatively describe $\eta/s$ and $\sigma/T$ above $T_C$, and a further development of the critical contribution to the transport coefficients in the critical region is needed. Recently a generic extension of hydrodynamics by a parametrically slow mode or modes (“Hydro+”) and a description of fluctuations out of equilibrium have been considered in Ref. \cite{Stephanov:2017ghc}.

\section{\label{sec4}Conclusion}

We have calculated the specific shear viscosity $\eta/s$ and the ratio of electric conductivity to temperature $\sigma_0/T$ of  QGP matter in the extended PNJL model for a wide range of quark chemical potentials using the framework of the Boltzmann equation in the relaxation time approximation. 

$\bullet$ We showed that both, the specific shear viscosity $\eta/s$ and the ratio of the electric conductivity to the temperature, $\sigma_0/T$,  depend strongly on the chemical potential.

$\bullet$ We demonstrated  the dependence of the transport coefficients on the quark relaxation times, which were estimated within two methods:
 either by using the averaged transition rates $\bar w_{ij}$  or by the 'weighted' thermal averaged cross sections $\bar \sigma_{ij}$. The evaluation made within the first method is considered to be more realistic as it stems from the derivation of the relaxation time through the interaction rate. 

$\bullet$ In the vicinity of the chiral phase transition both methods result in a similar temperature dependence of the considered transport coefficients, which are for a vanishing quark chemical potential in agreement with various results from the literature. They include the results for the specific shear viscosity $\eta/s$ and the ratio of the electric conductivity and the temperature, $\sigma_0/T$, obtained with the $N_f = 3$ NJL model \cite{Marty:NJL13}, lattice QCD predictions, the $N_f =2$ linear sigma model \cite{Heffernan:2020zcf}, predictions from the transport models such as the URQMD, the BAMPS, the SMASH, the PHSD and estimations from the dynamical quasiparticle model \cite{Soloveva:2019xph}. In the vicinity of the pseudo-critical temperature our results are remarkably close to that of lQCD calculations and to the results from the DQPM.
 
$\bullet$ The key result of this paper is the quark chemical potential $\mu_q$ dependence of transport coefficients. At  moderate values of $\mu_q$ ($\mu_q \leq 0.3$ GeV), where the phase transition is a rapid crossover, transport coefficients show a smooth temperature dependence while approaching the (pseudo)critical temperature from the high temperature region. 

$\bullet$ At large values of $\mu_q$ the presence of a first order phase transition  changes the temperature dependence of the transport coefficients drastically and a discontinuity can be seen when approaching the critical temperature.

$\bullet$ We found that the influence of the CEP on the evaluated transport coefficients is rather small in comparison to the modification due to a 1st order phase transition. 
For the specific shear viscosity a similar behaviour near the chiral phase transition has been obtained in the $N_f=2$ NJL model in Ref.~\cite{Sasaki}. 

$\bullet$ We have considered furthermore the dimensionless ratio of specific shear viscosity to the scaled electric conductivity.
It shows as well a discontinuity at $T_C$  if the chiral transition is  of 1st order but is otherwise almost constant for $T>2T_C$.

To wind up, we have found a significant dependence of the value of the considered transport coefficients on the quark relaxation times evaluated by two different methods, which can explain the difference in the previously known RTA results from other models. Nevertheless, in the vicinity of the chiral phase transition the temperature and chemical potential dependence of the transport coefficients is similar for the both presented methods.


\begin{acknowledgments}
The authors thank Wolfgang Cassing, Rudy Marty,  Taesoo Song,  and Juan Torres-Rincon for useful discussions.
O.S. acknowledge support from the \textquotedblleft Helmholtz Graduate School for Heavy Ion research\textquotedblright. 
O.S. and E.B. acknowledge support by the Deutsche Forschungsgemeinschaft (DFG, German Research Foundation) through the grant CRC-TR 211 'Strong-interaction matter under extreme conditions' - Project number 315477589 - TRR 211. 
This work is supported by the European Union’s Horizon 2020 research and innovation program under grant agreement No 824093 (STRONG-2020) 
and by the COST Action THOR, CA15213.  
\end{acknowledgments}



\end{document}